\documentclass[fleqn,usenatbib]{mnras}

\usepackage[T1]{fontenc}
\usepackage{ae,aecompl}

\usepackage{graphicx}	
\usepackage{amsmath}	
\usepackage{amssymb}	
\graphicspath{{figure/}} 
\usepackage{lipsum}
\usepackage{booktabs}
\usepackage{parskip}
\usepackage{dirtytalk}
\usepackage{placeins}
\usepackage{rotating}
\usepackage{lscape}
\usepackage{pdflscape}
\usepackage{multicol}
\usepackage{multirow}
\usepackage{mathtools}
\usepackage{enumerate}
\usepackage{blindtext}
\usepackage{bm}
\usepackage{amsbsy}
\usepackage{fixmath}
\usepackage{arydshln}
\usepackage[english]{babel}
\usepackage{float}
\usepackage[font=Large]{caption}
\usepackage{gensymb}
\usepackage[utf8]{inputenc}
\usepackage{fixltx2e}
\usepackage{epstopdf}
\usepackage{mwe}
\AppendGraphicsExtensions{.tiff}
\usepackage{subcaption}
\def\simgt{\lower.5ex\hbox{$\; \buildrel > \over \sim \;$}}

\def\aap{A\&A}
\def\apj{ApJ}
\def\mnras{MNRAS}
\def\apjs{ApJS}

\title[SEAGLE-I: Numerical Simulations of Strong Lenses]{SEAGLE -- I: A pipeline for simulating and modeling strong lenses from cosmological hydrodynamic simulations}

\author[Mukherjee et al.]{Sampath Mukherjee$^{1}$\thanks{\href{mailto:sampath@astro.rug.nl}{\nolinkurl{sampath@astro.rug.nl}}},
L\'{e}on V. E. Koopmans$^{1}$, R. Benton Metcalf$^{2,3}$, 
\newauthor
Nicolas Tessore$^{4}$, Crescenzo Tortora$^{1}$,
Matthieu Schaller$^{5}$, Joop Schaye$^{5}$, 
\newauthor
Robert A. Crain$^{6}$, Georgios Vernardos$^{1}$, Fabio Bellagamba$^{2,3}$, Tom Theuns$^{7}$
\\
\\
\\
$^{1}$Kapteyn Astronomical Institute, University of Groningen, PO Box 800, 9700AV Groningen, The Netherlands\\
$^{2}$Dipartimento di Fisica e Astronomia, Universit\`a di Bologna, via Gobetti 93/2, I-40129 Bologna, Italy\\
$^{3}$Istituto Nazionale di Astrofisica (INAF) - Osservatorio di Astrofisica e Scienza dello Spazio di Bologna, via Gobetti 93/3, I-40129 Bologna, Italy\\
$^{4}$Jodrell Bank Centre for Astrophysics, University of Manchester, Alan Turing Building, Manchester M13 9PL, UK\\
$^{5}$Leiden Observatory, Leiden University, PO Box 9513, 2300 RA Leiden, The Netherlands\\
$^{6}$Astrophysics Research Institute, Liverpool John Moores University, 146 Brownlow Hill, Liverpool, L3 5RF\\
$^{7}$Institute for Computational Cosmology, Department of Physics, University of Durham, South Road, Durham, DH1 3LE\\
}

\date{Accepted XXX. Received YYY; in original form ZZZ}

\pubyear{2017}

\begin{document}
\label{firstpage}
\pagerange{\pageref{firstpage}--\pageref{lastpage}}
\maketitle


\begin{abstract}
In this paper we introduce the {\it SEAGLE} (i.e.\ Simulating EAGLE LEnses) program, that approaches the study of galaxy formation through strong gravitational lensing,  using a suite of high-resolution hydrodynamic simulations, Evolution and Assembly of GaLaxies and their Environments (EAGLE) project. We introduce the simulation and analysis pipeline and present the first set of results from our analysis of early-type galaxies. We identify and extract an ensemble of simulated lens galaxies and use the {\tt GLAMER} ray-tracing  lensing code to create mock lenses similar to those observed in the SLACS and SL2S surveys, using a range of source parameters and galaxy orientations, including observational effects such as the Point-Spread-Function (PSF), pixelization and noise levels, representative of single-orbit observations with the Hubble Space Telescope (HST) using the ACS-F814W filter. We subsequently model these mock lenses using the code {\tt LENSED}, treating them in the same way as observed lenses. We also estimate the mass model parameters directly from the projected surface mass density of the simulated galaxy, using an identical mass model family. We perform a three-way comparison of all the measured quantities with real lenses. We find the average total density slope of EAGLE lenses, $t=2.26\; (0.25\; \rm{rms})$ to be higher than SL2S, $t=2.16$ or SLACS, $t=2.08$. We find a very strong correlation between the external shear ($\gamma$) and the complex ellipticity ($\epsilon$), with $\gamma \sim \epsilon/4$. This correlation indicates a degeneracy in the lens mass modeling. We also see a dispersion between lens modeling and direct fitting results, indicating systematical biases.
\\    
\end{abstract}



\section{Introduction}\label{introduction}

Massive early-type galaxies (ETGs) are expected to form during the later stages in the hierarchical formation process (\citealt{blumenthal1984,frenk1985}). ETGs in the local universe follow a number of well-known relations or correlations between their velocity dispersion, stellar age, chemical composition (\citealt{bender1992, bender1993}), and exhibit a small scatter around the nearly-isothermal central density profiles (e.g., \citealt{rusin2003a, rusin2003b, rusin2005, koopmans2006, koopmans2009}). Galaxy formation models are only now beginning to address the origin of these empirical scaling relations accounting for the physical processes that play a role in their formation. There are various possibilities for their formation, for example via monolithic collapse (\citealt{eggen1965, searle1978}), mergers of lower-mass (disk) galaxies (\citealt{toomre1972, schweizer1982}), satellite accretion (\citealt{searle1978}), and hierarchical merging (\citealt{white1978, fall1983}). Various environmentally dependent evolutionary processes such as stripping (\citealt{gunn1972}), cannibalism (\citealt{ostriker1977}), stretching (\citealt{barnes1992}), harassment (\citealt{moore1998}), strangulation (\citealt{balogh2000}), squelching (\citealt{tully2002}) and splash-back (\citealt{fukugita2006}) have been proposed to explain the formation-mechanisms of early-type galaxies. The explicit study of their structure (\citealt{navarro1996, moore1998}), formation and subsequent evolution provides a powerful test of the (dis)agreement between observations and the $\rm \Lambda$CDM paradigm.

\cite{loeb2003} suggest that the inner regions might behave as dynamical attractors, whose phase-space density is nearly invariant under the accretion of collisionless matter (\citealt{gao2004, kazantzidis2006}). In this scenario, one might expect less structural evolution of the inner regions of massive early-type galaxies at $z < 1$, compared to models in which most gas had not yet turned into stars before the mass assembly of their inner regions took place. Hence, one way to study the formation scenario of massive elliptical galaxies is to quantify the evolution of the mass distribution in their inner regions in the redshift range $0<z<1$.

Over the last few decades, tremendous progress has been made in our understanding of cosmic structure and galaxy formations mechanisms. This is in part due to (semi) analytic galaxy-formation theory giving us detailed calculations of the Cold Dark Matter (CDM) power spectrum (\citealt{peebles1982,blumenthal1984}), Press-Schechter theory (\citealt{press1974}), the statistics of peaks in Gaussian Random fields (\citealt{bardeen1986}) and galaxy formation models (\citealt{white1978}). Analytical approaches have their limitations though in addressing more complicated physical processes. In the absence of precise analytical methods for computing for example the non-linear dark matter power spectrum, the properties of dark matter substructure, etc., full-scale numerical simulations are the only method available. Semi-numerical models have also been employed, building on numerical simulations. The combined results of these semi-analytical and numerical simulations have provided valuable insight into the study of galaxy formation over the last two decades  (\citealt{frenk1999,springel2005b,springel2005a, springel2006,springel2010,schaye2010,vogel2012,vogel2014,s15}). 

Strong gravitational lensing due to ETG provides a robust observational test of a number of theoretical predictions for galaxies at $z \leq 1$, especially when it is being combined with stellar kinematic data (\citealt{treu2004, sand2004, koopmans2006}). Employing the results of the lens models (\citealt{koopmans2006}), some studies (\citealt{treu2006}) quantified the degree of homogeneity in the inner density profiles of the early-type galaxies, suggesting close to isothermal density profiles on average, but with a scatter. Many questions however remain unanswered. To study strong-lensing ETGs in more detail, the Sloan Lens ACS Survey (SLACS, \citealt{bolton2006,treu2006,koopmans2006,gavazzi2007,bolton2008a,gavazzi2008,treu2009,auger2009,auger2010a,auger2010b,newton2011,shu2015,Shu2017}) and the Strong Lensing Legacy Survey (SL2S, \citealt{ruff2011,gavazzi2012,sonnenfeld2013a,sonnenfeld2013b,sonnenfeld2015}) have provided relatively uniform samples. The  lens models from these surveys, however, have not yet been compared in detail to high resolution numerical simulations. An exception is \cite{Bellagamba2017} who found a significantly shallower slope for the dark matter alone by preforming a detailed study of one lens.
 
In this paper, we present a new lens-galaxy simulation and analysis pipeline using the EAGLE simulations (\citealt{s15,c15}) and compare the results from mock lens projections to those of SLACS and SL2S. We introduce an automatic prescription that creates, models, and analyzes simulated lenses. We introduce a weighing scheme necessary to reduce the selection bias and statistically compare the simulated lenses with observations. We also probe the systematic errors and biases arising from the different line of sight projections and environmental effects. We find that using a simplex parameter estimator, we can quite robustly obtain the key lensing observables e.g., the Einstein radius, and mass density slope etc. We put-forward the concept of a 2D-complex space involving axis ratio and position angle in order to disentangle the degeneracy among them. 

The paper is structured as follows. In Section \ref{simulations and software} we summarize the EAGLE galaxy formation simulations and the relevant codes that we use in this paper. Section \ref{pipeline} describes the simulation and analysis pipeline that we have constructed. Lens modeling details are explained in Section \ref{lensmod}. The results of our mock-lens analyses are described in Section \ref{results}. We compare our mock-lens samples and their properties with observations in Section \ref{comparisons} and conclude with a summary in Section \ref{discussions}. Throughout the paper we use EAGLE simulations that assume a Chabrier stellar Initial Mass Function (IMF, \citealt{chabrier2003}). The values of the cosmological parameters are $\mathrm{\Omega_\Lambda}$ = 0.693, $\mathrm{\Omega_b}$ = 0.0482519, $\mathrm{\Omega_m}$ = 0.307, ${h=H_0/(100\; {\rm km\; s^{-1} \; Mpc^{-1}})}$ = 0.6777 and ${\sigma _{8}}$ = 0.8288. These are taken from the Planck satellite data release (\citealt{planck2014}), again in agreement with the EAGLE simulations.

\begin{table*}
 \label{simtable}
 \begin{tabular}{c c c c c c c} 
 \hline
 Name & L (cMpc) & N & $m_{\rm g}$ ($\rm M_{\rm \odot}$) & $m_{\rm DM}$ ($\rm M_{\rm \odot}$)& $\epsilon_{\rm com}$ (ckpc)& $\epsilon_{\rm prop}$ (pkpc)\\ [0.5ex] 
 \hline
 L025N0376 & 25 & $376^3$ & $1.81 \times 10^6$ & $9.70 \times 10^6$ & 2.66 & 0.70 \\ 
 L025N0752 & 25 & $752^3$  & $2.26 \times 10^5$ &  $1.21 \times 10^6$ & 1.33& 0.35\\
 L050N0752 & 50 & $752^3$  & $1.81 \times 10^6$&$9.70 \times 10^6$ & 2.66 & 0.70\\
 L100N1504 & 100 & $1504^3$  & $1.81 \times 10^6$ &$9.70 \times 10^6$ & 2.66 & 0.70\\ [1ex]
 \hline
\end{tabular}
\caption{\normalsize The main EAGLE simulations. From left to right: simulation name suffix; comoving box size; number of Dark Matter (DM) particles (initially an equal number of baryonic particles are present); initial baryonic particle mass; DM particle mass; comoving Plummer-equivalent gravitational softening length; maximum proper softening length (reproduced from \citealt{s15}). Throughout the paper proper kpc is used synonyously with kpc unless otherwise mentioned.}
\label{simtable}
\end{table*}

\section{Numerical Codes}\label{simulations and software}

\begin{figure*}
\includegraphics[width=\textwidth]{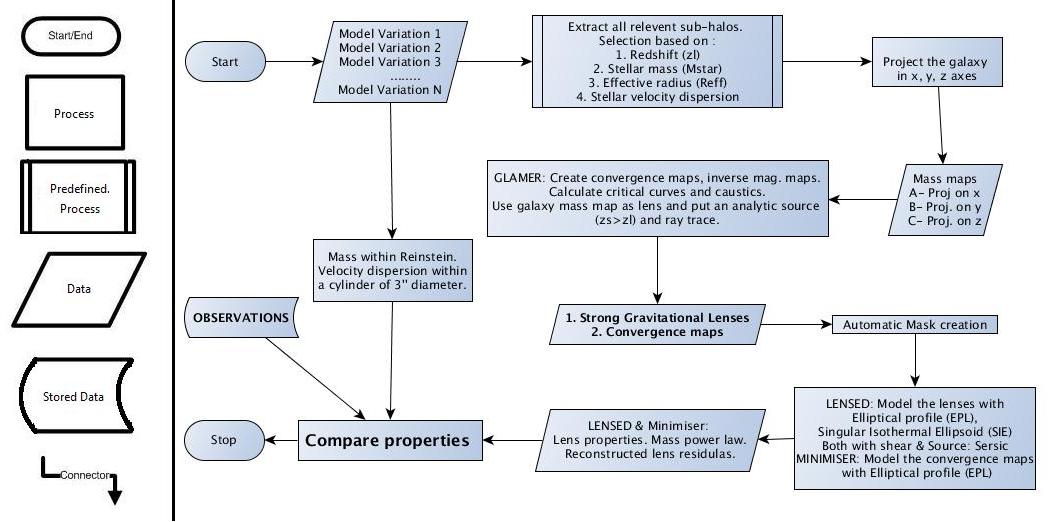}
\caption{\normalsize The SEAGLE flow chart showing that the convergence mass maps -- simulated using GLAMER and galaxies extracted from EAGLE -- are analyzed  via two different channels, i.e.\ via the modeling of the simulated lensed images, and via direct fitting of the same (lens) mass model to the convergence mass map. The two resulting parameter sets are compared to each other and to the corresponding observables coming from the SLACS and SL2S surveys.}
\label{flowchart}
\end{figure*}

In this section we briefly describe the simulations, numerical codes, and tools that are used in this work. We describe the EAGLE hydrodynamical simulations from which we select lens galaxies (Section~\ref{eagle}), the {\tt GLAMER} ray-tracing code to simulate mock lenses for various lens orientations and sources (Section~\ref{glamer}), and the {\tt LENSED} lens-modeling code used to infer mass-model parameters (Section~\ref{lensed}). 

\subsection{Galaxy-Formation Simulations from EAGLE}\label{eagle}

Evolution and Assembly of GaLaxies and their Environment (EAGLE)\footnote{\url{http://icc.dur.ac.uk/Eagle/}} is a suite of hydrodynamical simulations of the formation of galaxies and super-massive black holes in a $\rm \Lambda$CDM universe (\citealt{s15,c15,McAlpine2016}). EAGLE simulations are carried out using the modified N-Body Tree-PM (Particle Mesh) SPH (Smoothed Particle Hydrodynamics) code {\tt GADGET 3} (\citealt{springel2005a}). The resulting galaxies are in good agreement with observations of the star formation rate, passive fraction, Tully-Fischer relation, total stellar luminosity of galaxy cluster and colors (\citealt{s15,trayford2015}), the evolution of the galaxy stellar mass function and sizes (\citealt{furlong2015b,furlong2015a}), rotation curves (\citealt{schaller2015a}) and the $\alpha$-enhancement of ETGs (\citealt{Segers2016}). \\ 
\ \\
The subgrid physics employed in EAGLE is based on that developed for OWLS (\citealt{schaye2010}) and used also in {\tt GIMIC} (\citealt{crain2009}) and cosmo-OWLS (\citealt{brun2014}). The modifications to the SPH implementation together are known as `Anarchy' (\citealt{mschaller2015}). EAGLE galaxies are defined as gravitationally bound sub-halos identified by the {\tt subfind} algorithm (\citealt{springel2001, dolag2009b}). The gravitational softening is summarized in Table~\ref{simtable}. 

In this paper we have chosen to use the {\tt Reference} model having L050N0752 (see Table \ref{simtable} and \citealt{s15,c15}) to ensure that we have a coherent sample of galaxies that have been formed from an identical set of initial conditions subjected to different physical models when comparing results between different model variations of EAGLE (\citealt{c15}) in forthcoming papers. For detailed descriptions on the various galaxy-formation prescriptions and sub-grid physics we refer to \citet{s15} and \cite{c15}.

\begin{figure*}
\minipage{0.326\textwidth}
  \includegraphics[height=\textwidth, width=\textwidth]{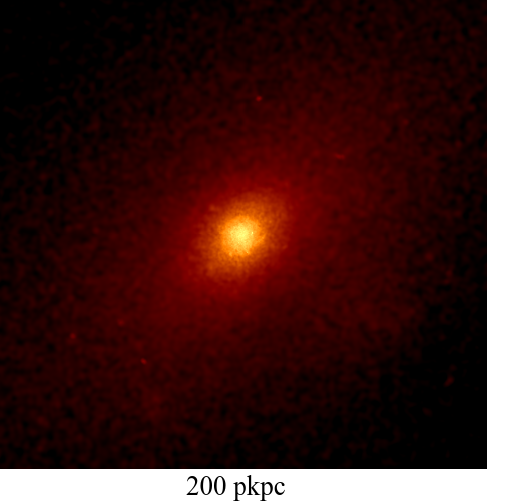}
\endminipage
\minipage{0.326\textwidth}
  \includegraphics[height=\textwidth, width=\textwidth]{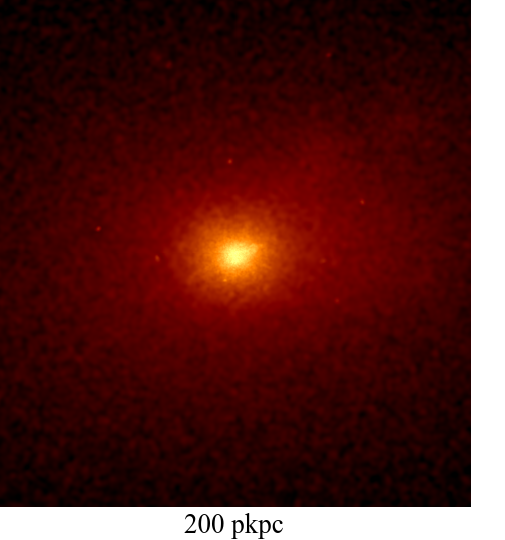}
 \endminipage
\minipage{0.326\textwidth}%
  \includegraphics[height=\textwidth, width=\textwidth]{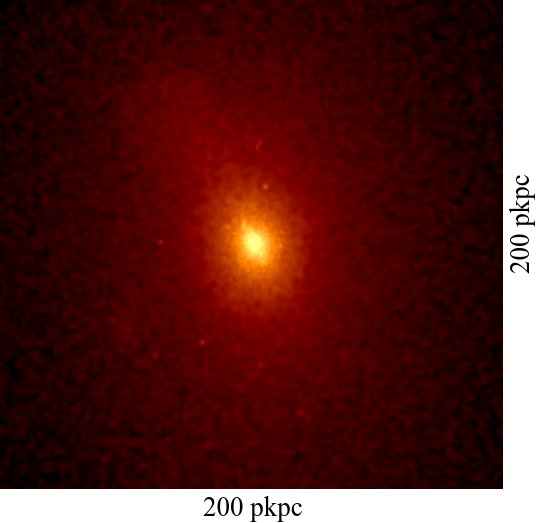} 
  \endminipage 
\caption{\normalsize An example of mass maps of a typical ETG of stellar mass =$1.9 \times 10^{11} {\rm \;M_{\bigodot}}$ at $z=0.271$, extracted from the Reference (L050N0752) EAGLE simulation and the box-size is 200 pkpc. Frame [left] is the visualization of projected mass map of the galaxy having axis ratio, q $\approx 0.76$ when the line of sight is rotated by (90, 0, 0) $\deg$ i.e., 90 $\deg$ rotation in  $x$ axes with respect to the center of the simulation box. Frame [middle] displays the galaxy having q $\approx 0.72$ when our focus has been rotated by (0, 90, 0) $\deg$ i.e., 90 $\deg$ rotation in  $y$ axes. Frame [right] displays the galaxy with q $\approx 0.69$ when the rotation angle is (0, 0, 90) degrees in $z$ axes.}
\label{los}
\end{figure*}

\subsection{Strong Lens Simulations with GLAMER}\label{glamer}

{\tt GLAMER}\footnote{\url{http://glenco.github.io/glamer/}} is a ray-tracing code for the simulation of gravitational lenses (\citealt{metcalf2014,petkova2014}). The deflection angles, shear, and other relevant properties are calculated using a modified tree algorithm described in \cite{barnes1989}. It uses Adaptive Mesh Refinement (AMR) in ray-casting, based on the requirements of the source size, location and surface brightness distribution and to find critical curves and caustics. Ray paths are determined from the observer to the source plane through multi-plane deflection, convergence, and shear calculations. {\tt GLAMER} allows for a wide variety of source types and the mass distribution on each lens plane can be represented in several different ways, for example via a surface density map in FITS format. The resulting lensed images are subsequently convolved with a point spread function (PSF) and appropriate noise levels can be added. For further details one can consult GLAMER I \& II papers (\citealt{metcalf2014,petkova2014}).

In this paper, we use a single lens plane for representing the convergence of galaxies extracted from EAGLE, because the maximum box size ($<100$\,Mpc) is still small compared to the cosmological distances involved. This can be expanded to multiple lens planes for much larger boxes.  We also assume an elliptical S\'{e}rsic profile for the sources with varying parameters, placed inside the diamond caustic to generate preferentially highly magnified systems, similar to those found in the SLACS and SL2S surveys. All of these choices can be varied in the pipeline if desired.

\subsection{Gravitational Lens Modeling with LENSED}\label{lensed}

{\tt LENSED}\footnote{\url{http://glenco.github.io/lensed/}}  is a publicly available code which performs parametric modeling of strong lenses by taking advantage of the massively parallel ray-tracing kernel on a graphics processing unit (\citealt{tessore15a}) to perform all necessary calculations. Combining these accurate and fast forward simulations with the Nested-Sampling Bayesian analysis, {\tt MULTINEST} (\citealt{feroz2008,feroz2009, feroz2013}), allows the simultaneous optimization of tens of non-linear parameters of the selected model (e.g. lens and source) and the full posterior probability distribution for the mass distribution and the background source in typically 10 minutes on a multi-core machine. 
The setup of the physical system, priors, input files including images, masks, PSFs and noise maps can be done  using a single configuration file. The code reports statistically well-justified errors, including degeneracies, for the lens model parameters i.e., the full posterior reconstruction, and also simultaneous fitting of sources and lenses. {\tt LENSED} has been well tested on SLACS lenses. For details see \cite{tessore15a} and \cite{Bellagamba2017}.

\begin{table*}
\label{properties}
\begin{center}
\begin{tabular}{l l l } 
\hline
\hline
\multicolumn{3}{c}{\bf Galaxy Selection}\\
Observable & Value & Comments\\
\hline
$\rm M_{\star}$ & $\rm \geq 1.76 \times 10^{10} M_{\bigodot}$ & Stellar mass lower threshold. Taken from \cite{auger2010a} \\
$\sigma$ & > 120 km/sec& Stellar velocity dispersions are kept lower than SLACS\\
$\mathrm{R_{50}}$ &> 1 kpc & Half mass projected radius\\ \hdashline
$\rm M_{\star}$&$\rm >10^{11}\;M_{\bigodot}$& Stellar mass lower cut-off for comparison with observations\\
\hline
\multicolumn{3}{c}{\bf Lens Candidates}\\
Object-properties & Value & Comments\\
\hline
Sim. used & Reference (L050N0752)& 50 cMpc box is best for comparing with other scenarios\\
Orientation& x, y and z axis & Projected surface density maps are made along each axis\\
Redshift & ${\rm z_{l}=0.271}$& Consistent with SLACS' mean lens-redshift of 0.3  \\
No. of galaxies & 252& Total number of galaxies satisfying our selection criteria\\
No. of projected galaxies& 756& Total number of galaxies after projection on 3 axes\\
\hline
\multicolumn{3}{c}{\bf Source Properties}\\
Parameters & Value & Comments\\
\hline
Source Type& S\'{e}rsic& Consistent with analyzed SLACS lenses (\citealt{newton2011})\\
Brightness&23 apparent mag.& \Large{$''$}\\
Size ($R_{\rm eff}$)& 0.2 arcsec& \Large{$''$}\\
Axis ratio ($q_s$)& 0.6& \Large{$''$}\\
S\'{e}rsic Index & 1 & \Large{$''$}\\
Redshift&$\rm z_{s}$=0.6& \Large{$''$}\\
Position & Random within caustics& Producing more rings and arcs lens systems, consistent with SLACS\\
\hline
\multicolumn{3}{c}{\bf Instrumental Settings}\\
Parameters & Value & Comments\\
\hline
PSF &  Gaussian, FWHM=0.1 arcsec& -\\
Noise & HST ACS-F814W, 2400 sec & -\\
\hline
\multicolumn{3}{c}{\bf Image Properties}\\
Map used & Properties & Value\\
\hline 
\multirow{2}{*}{Surface density}& (a) Size & 512$\times$512 pixels\\
& (b) Units&kpc\\
\multirow{2}{*}{$\kappa$, Inv. mag. map and Lens}& (a) Size & 161$\times$161 pixels\\
& (b) Units&degrees (converted from arcsec)\\
\hline
\hline
\end{tabular}
\label{properties}
\end{center}
\caption{\normalsize The summary of the current SEAGLE pipeline settings.}
\label{properties}
\end{table*}

\section{Pipeline}\label{pipeline}

In this section we describe the SEAGLE (Simulating EAGLE LEnses) pipeline in more details. We describe the selection criteria of the lens candidates from EAGLE (Section \ref{twins}), the lens galaxy extraction technique (Section \ref{extracting}), the line-of-sight projection effects on the shape of lens galaxies (Section \ref{losprojection}), the method to create mock lens systems with {\tt GLAMER} (Section \ref{lenscre}), the automatic mask creation process (Section \ref{mask}), and details of the final lens sample used in this paper (Section \ref{lenssampleI}). The simulation and analysis pipeline is shown in Figure \ref{flowchart}. \\

In this paper extraction of galaxies is done at one particular redshift and the resulting mass distribution is projected along one of the three principal axes of the simulation box. A \cite{sersic1968} source is then placed at a random source position within the diamond caustics at a higher redshift as experimented with {\tt GLAMER}. The PSF and noise are similar to those for a single orbit HST ACS-F814W observations to make the mock lenses appear similar to observed SLACS lenses. The resulting lenses are subsequently modeled and  analyzed by comparing their ensemble properties with those from SLACS and SL2S. We note that most of the above choices can be easily modified. 

\begin{figure*}
\begin{center}
\includegraphics[width=\textwidth]{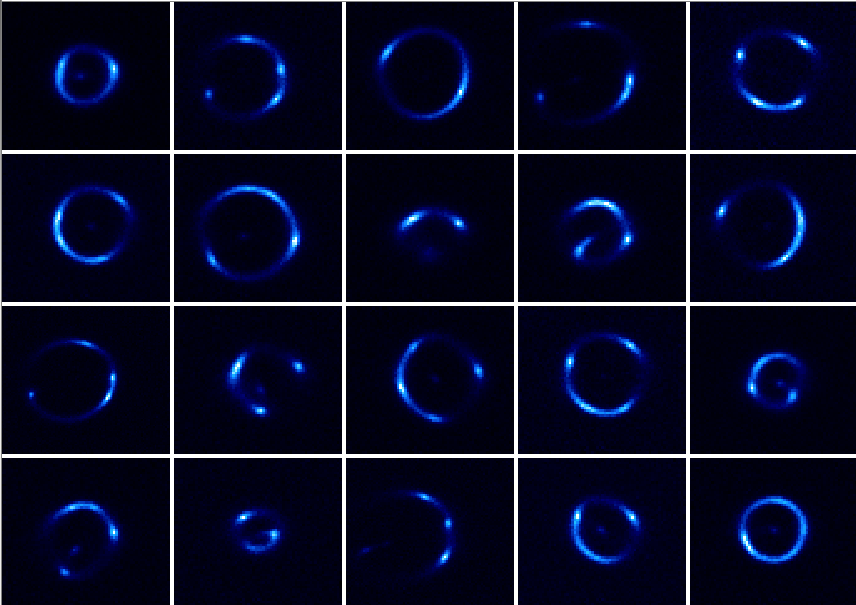}
\end{center}
\caption{\normalsize A subset of strong lenses from EAGLE (Reference model) 50 cMpc, $z_l =0.271$. Even some of the rare SLACS lenses has been mimicked very well via our pipeline. The sub kpc fluctuations however cannot be simulated with a simple S\'{e}rsic source object. But that is not necessary for having a statistical sample of simulated SLACS like lenses. We have not put the lensing galaxy in the foreground so there is no contamination of the light from the foreground ETG.}
\label{slacsvseagle}
\end{figure*}

\subsection{Lens-Galaxy Selection}\label{twins}

The next generation of lens surveys (for example with {\sl Euclid} \citep{lau2011}) are expected to increase the number of lenses by orders of magnitude, in particular finding lower mass and smaller image separation lenses. This will increase the parameter space of strong lenses in terms of their mass, stellar velocity dispersion and other observables considerably. The selection criteria for extracting galaxies from EAGLE, however, are based on parameters obtained from currently confirmed strong lenses due to ETGs, in particular from SLACS. Keeping this restriction in mind, we explore a volume-limited sample of lens galaxies with observables (e.g.\ stellar mass) in the range of ETGs from SLACS (\citealt{auger2009, auger2010a,auger2010b}) and SL2S (\citealt{sonnenfeld2013a,sonnenfeld2013b}). The SLACS sample consists of a wide ranges of photometric and spectroscopic measurements using HST and SDSS and inferred data products, which include for example, the parameters inferred from lens modeling and stellar-population analysis \citep{auger2010b}. The parameter space of SLACS broadly overlaps with SL2S lenses, which makes it useful to compare properties of simulated lenses with both samples.  

The initial selection is based on lens redshift ($z_{\rm l}$) and stellar mass ($\rm M_{\star}$) in accordance with \citet{auger2010a}, where the lens redshift range is $ 0.1 \leq z_{\rm lens} \leq 0.3$ and the stellar mass threshold is $\rm M_{\star} \geq 1.76 \times 10^{10} \;M_{\bigodot}$. No upper limit is set. The stellar velocity dispersion ($\sigma$) and half mass radii ($\mathrm{R_{50}}$), which is a proxy for effective radii ($\mathrm{R_{eff}}$) in observations, are only used to clip outliers e.g.,\ due to halo stars, mergers and other contaminations arising from stray particles in the simulations. Table \ref{properties} summarizes the details of our selection criteria. 

We find it difficult to implement an automated recipe for the lens modeling for galaxies with stellar masses, $ \rm M_{\star} < 10^{11}\; M_{\bigodot}$. This is due to the resolution effect of the particles during projection, which creates prominent but artificial images in the central regions of the lenses after ray tracing, which are not seen in real lens galaxies. In order to implement an automated lens modeling scheme with {\tt LENSED} we therefore further restrict ourselves to galaxies with $ \rm M_{\star} > 10^{11}\; M_{\bigodot}$ (calculated within a cylinder of $1.5^{\prime \prime}$ in radius, consistent with SLACS) which produce extended arcs and rings (see Figure~\ref{slacsvseagle}). These are far less affected by any resolution effect and are still within the upper mass range of SLACS and SL2S lenses. To down-weight lower-mass galaxies in the volume limited set of EAGLE lenses in comparisons to SLACS or SL2S, in Section \ref{weightscheme} we introduce a weighting scheme based on their lensing cross-section, that compensates for the observational selection biases and allows for a more accurate comparison between the simulations and the observations.
We ignore the magnification bias, which we assume to vary more slowly with galaxy mass unlike the cross section.

\subsection{Galaxy/Halo Extraction}\label{extracting}

To extract a galaxy from the EAGLE snapshots we use the {\tt Friends-Of-Friend (FoF)} catalogs. We use the stellar mass catalog from the snapshots and particle data at the desired redshift of the currently used {\tt Reference} simulation (i.e., L050N0752). The choice of aperture is important given its direct effect on the stellar mass calculation for massive and extended galaxies with $\rm M_{*} > 10^{11} M_{\bigodot}$ (see \citealt{s15}). Given that most lens-galaxies have half-mass radii of 5-10 kpc, we choose a 10-kpc aperture to select the closest analogues to observed lenses from the simulations. We select all sub-halo indices that match our selection criteria, and reject any galaxy having half-light radii < 1 kpc in the EAGLE catalogs (these objects are misidentified galaxies and lie far from the Fundamental Plane). This aperture size avoids inclusion of spurious stellar mass which would be discounted in the modeling of observed lens galaxies using e.g. a smooth S\'{e}rsic profile. Eagle catalogs have {\tt GroupNumber} and {\tt SubGroupNumber} which are numbers assigned to {\tt FoF} group and subgroup respectively. They are numbered according to their decreasing masses. That means subgroup 0 of a {\tt FoF} group corresponds to the most massive subgroup within the group. We read the {\tt GroupNumber} and {\tt SubGroupNumber} using the same indices to recover the {\tt FoF} Group ID and {\tt Subfind} subgroup ID and subsequently select all their particles and obtain their meta-data from the simulations, using the group IDs. Galaxy selection and outlier rejection are currently fully automated in the pipeline and the criteria can be altered if necessary.

\begin{figure*}
\includegraphics[width=\textwidth]{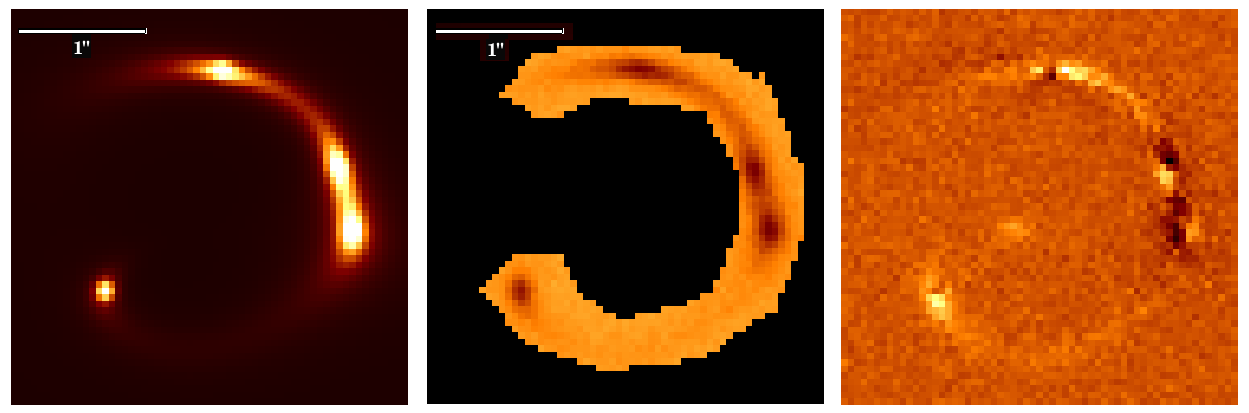}
\caption{\normalsize The left panel shows an example of a simulated lens with noise and PSF (see Table~\ref{properties} for details). The middle panel shows the reconstructed image of the lens inside the mask, using {\tt LENSED}. The right panel shows the unnormalized residual image of the data minus the model. The peak brightness of simulated lens and residuals are 1.30 and 0.37, respectively.}
\label{gullu}
\end{figure*}

\subsection{Line-of-sight Projection}\label{losprojection}

Once the catalogs of dark-matter, stellar, gas and black hole particles of a galaxy have been extracted from the simulations, we allow for any arbitrary spatial rotation. We rotate particle position vectors around the center of the lens galaxy. Although this does not lead to an independent lens galaxy, it does allow for some testing of the effects of orientation on the inference of the galaxy properties. 

Figure \ref{los} shows how the projected shape of a galaxy changes when viewing it from three different angles. In this paper we use each galaxy three times, projected along each of the three principle axes of the simulation box. In the future papers we use these to assess systematics due to projection of the main galaxy halo and line-of-sight effects in the nearby environment of the lens (i.e.\ inside the box). The particles are then converted into projected mass maps after smoothing of the particles with the same SPH kernel as used in the simulation (for details see appendix A of \citealt{trayford2017}). 

\begin{figure}
\includegraphics[width=\columnwidth]{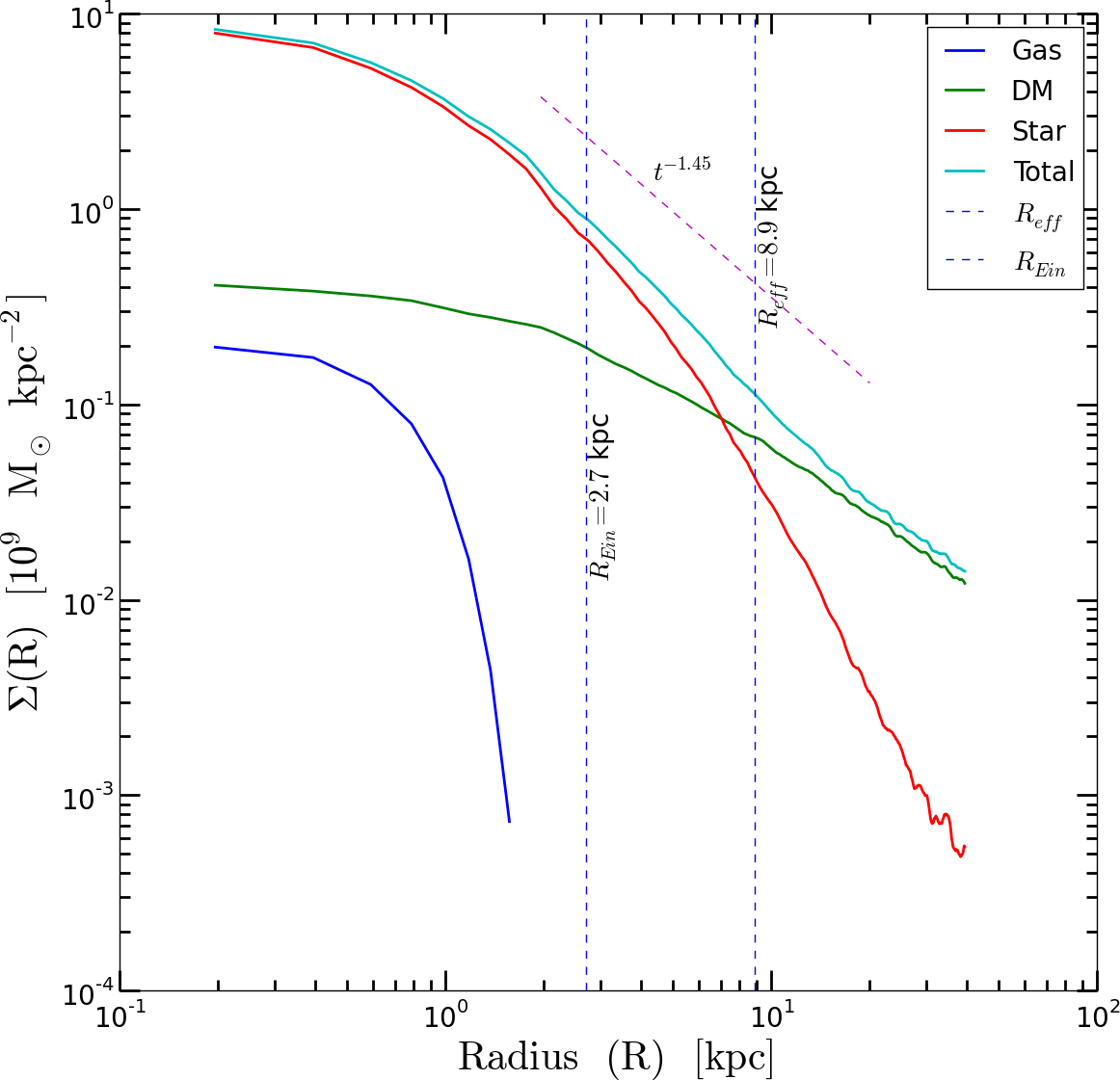}
\caption{\normalsize Surface density profiles of DM, stars, gas and the total mass of a typical ETG from EAGLE. The effective radius of the galaxy and the slope of best fitted mass model profile along with Einstein radius, as obtained from lens modeling, are also indicated.}
\label{auto}
\end{figure}

We also simultaneously calculate the surface density profiles of the matter distribution for each projected mass map. The surface densities for individual particle types (DM, stars and gas) and a total mass profile are calculated separately. Figure \ref{auto} shows a typical example for an ETG's mass profiles. The effect of the resolution of the simulation inside $\sim$1 kpc is clearly visible.

The resolution of the simulation plays a role in the core of the galaxy, where we hit the resolution limit (see Appendix \ref{Eapp}). 2-3 times the gravitational softening length, which is independent of the density, away from the core its effect no longer plays a crucial role. So we mask the central pixels in the lensed images. In subsection ~\ref{mask} we describe this in details.

\subsection{Mock Lens-System Creation}\label{lenscre}

The surface density maps are created on grids of 512 $\times$ 512 pixels (Table \ref{properties}), in units of solar mass per pixel, and form the input to {\tt GLAMER}. The width (100 pkpc) and pixel scale (0.2 pkpc) of the grid ensure the surface density map and corresponding convergence map are well-resolved in the relevant regions (see \citealt{Tagore2018}), down to the softening length and consistent with SLACS resolution of 0.05 arcsec (at  $z$=0.271, SLACS resolution corresponds to $\approx$ 0.2 pkpc). We then choose a lens and source redshifts for {\tt GLAMER} to convert these mass maps into convergence maps. For each mass map, the critical curves and caustics are calculated to determine where a source has to be placed in order to create multiple lensed images.   
In this paper we use an elliptical S\'{e}rsic brightness profile of the source with index $n=1$. Its apparent magnitude is constant at 23 in the HST-ACS F814W filter (AB system) and its redshift is $z_{\rm s} =0.6$. The other parameters are its effective radius of 0.2 arcsec, a position angle $ \phi_s= 0 \deg$ and a constant axis ratio $q_s$=0.6. Given that source and galaxy position angles are uncorrelated, this fixed position angle of the source does not reduce generality. The source is placed randomly inside the diamond caustics of the lens. The pixel scale is  0.05\,arcsec, and the PSF and noise correspond to an HST-ACS-F814W  exposure of typically 2400\,s. The resulting images have a size of 161 $\times$ 161 pixels of 8.0 arcsec. The above parameter values are currently fixed for each lens, but are typical for the sample space of SLACS lenses \citep{koopmans2006,newton2011,bandara2013}. Since our goal is to assess global properties of the lenses, the precise choice of the source model (which is an exponential disk here; S\'{e}rsic with n=1) is currently of secondary importance. The images are exported in standard fits-file format. Table \ref{priortable} lists all parameter values. We like to point out that in this work, only arcs and rings lenses are simulated. No two image system are simulated since the number counts of two image system in SLACS is 6/84 $\sim$ 7\% (\citealt{auger2010a}) and 3/56 $\sim$ 6\% in SL2S (\citealt{sonnenfeld2013a}) are very low and no evidence of the lens properties being a function of lens geometry is reported (\citealt{auger2010b,sonnenfeld2013b}). So we assume non inclusion of the two image lenses are highly unlikely to bias the overall statistics.  

In addition to the simulated lenses, we store convergence maps and inverse magnitude maps of each lens galaxy. The brightness distribution of the lensing galaxy is not added to the lensed image grid. We assume that subtraction of the surface brightness distribution of the lens
galaxy can be done to sufficient accuracy that it does not affect the analysis in the current paper. Hence we assume little covariance between the source and lens brightness distributions. Experience with high-resolution HST-quality data of lenses in the I-band confirms this, although this assumption might not hold for lower-resolution ground based data. 

\subsection{Mask Creation}\label{mask}

The strong lens systems created using {\tt GLAMER} are modeled similarly to a real lens system. Masks are generated automatically in order to enable direct comparison with the lens models to the region around the lensed images. We do this for each lens by convolving the noisy lensed images with a Gaussian with a FWHM of 0.25 arcsec to reduce the noise and  smear the images to a slightly larger footprint. We then set a surface brightness threshold for the mask being a factor of typically 2.5--5 below the original noise. Pixels above the threshold are set to one and all others to zero. The mask then traces the surface brightness of the lensed images well below the noise level. We set the threshold values such that the mask bounds the significant surface brightness pixels but leaves a padding area that is largely noise example, middle panel of Figure \ref{gullu}. The central 7$\times$7 pixels are also masked to  remove any artificially bright central images as a result of the finite size of the SPH kernel and limited resolution. 
The final mask is used in the modeling and minimizer fitting in the following steps.

\begin{table*}
\caption{\normalsize The prior settings used in the lens-modeling with {\bf SIE + shear} and {\bf EPL + shear} mass model in {\tt LENSED}.}
\label{priortable}
\begin{center}
\begin{tabular}{l l l l l l l}
\hline
\hline
\multicolumn{7}{c}{\bf Priors used in ${\tt LENSED}^{\star}$}\\
\hline
\hline
\ \\
\multicolumn{7}{c}{\bf Elliptical Power Law (EPL) + Shear}\\
\hline
Parameter & Prior type${}^{\star \star}$ & \multicolumn{4}{c}{ Prior range}  & \multicolumn{1}{c}{Description}\\
& & $\mu $&$\sigma$& min & max & \\ 

\hline
$x_{\rm L}$      & norm &80.0 & 5.0 & -&- & Lens position: x coordinate\\                                           
$y_{\rm L}$    & norm &80.0 & 5.0 &- &- &Lens position: y coordinate\\                     
$r_{\rm L}$      & unif &- & -&5.0 & 70.0 & Einstein radius in pixel units\\
$t_{\rm L}$ & norm&1.1 &0.1&- &- & Surface mass density slope \\
$q_{\rm L}$     & unif &- &- &0.2 & 0.99& Lens axis ratio\\
${\phi}_{\rm L}$     & unif & -&- &0.0 & 180.0 &Lens position angle in degrees, wrapped around\\
${\gamma_1}_{\rm L}$     & norm& 0.0 & 0.01& -&- & Shear vector\\
${\gamma_2}_{\rm L}$     & norm &0.0 & 0.01& -& -& Shear vector\\
$x_{\rm S}$   & norm &80.0 & 30.0 & -& -& Source position: x coordinate\\                          
$y_{\rm S}$    & norm &80.0 & 30.0 &- &- & Source position: y coordinate\\                                            
$r_{\rm S}$    & unif &- & -&0.1 & 10.0 & Source size in pixel units\\
$mag_{\rm S}$  & unif &- &- &-5.0 & 0.0 & Source magnitude, adjusted with the background magnitude${}^\#$\\
$n_{\rm S}$    & norm &1.0 &  0.1 & -& -& S\'{e}rsic index \\
$q_{\rm S}$    & norm &0.5 & 0.1 & -&- & Source axis ratio \\
${\phi}_{\rm S}$   & unif & -& -&0.0 & 180.0 & Source position angle in degrees, wrapped around\\
\hline
\ \\
\multicolumn{7}{c}{\bf Singular Isothermal Ellipsoid (SIE) + Shear}\\
\hline
Parameter & Prior type${}^{\star \star}$ & \multicolumn{4}{c}{Prior range} & \multicolumn{1}{c}{Description}\\
& & $\mu $&$\sigma$& min & max & \\ 
\hline
$x_{\rm L}$      & norm &80.0 & 5.0 & - &- &Lens position: x coordinate\\                                           
$y_{\rm L}$    & norm &80.0 & 5.0 & - & -&Lens position: y coordinate\\                                         
$r_{\rm L}$      & unif &- & -&5.0 & 70.0  & Einstein radius in pixel units\\
$q_{\rm L}$     & unif &- & -&0.2 & 0.99& Lens axis ratio\\
${\phi}_{\rm L}$     & unif &- & -&0.0 & 180.0 & Lens position angle in degrees, wrapped around\\
${\gamma_1}_{\rm L}$     & unif& - & -&-0.1 & 0.1&Shear vector\\
${\gamma_2}_{\rm L}$     & unif &- & -&-0.1 &0.1 &Shear vector\\
$x_{\rm S}$   & norm &80.0 & 30.0 & - &- & Source position: x coordinate\\                          
$y_{\rm S}$    & norm &80.0 & 30.0 & - &- &Source position: y coordinate\\                                            
$r_{\rm S}$    & unif &- &- &0.1 & 10.0 & Source size in pixel units\\
$mag_{\rm S}$  & unif & -& -&-5.0 & 0.0 & Source magnitude, adjusted with the background magnitude${}^\#$\\
$n_{\rm S}$    & unif &- &- &0.5 &  2.0 & S\'{e}rsic index \\
$q_{\rm S}$    & unif &- & -&0.2 & 0.99 & Source axis ratio \\
${\phi}_{\rm S}$   & unif &- &- &0.0 & 180.0 & Source position angle in degrees, wrapped around\\
\hline
\multicolumn{7}{l}{$\star$ All values are in pixels except $q$, $\gamma$, $t_{\rm L}$, $mag_{\rm S}$, $n_{\rm S}$, and $\phi$. $\star \star$ norm = Gaussian (with mean $\mu$ and standard dev. $\sigma$), unif = Uniform}\\
\multicolumn{7}{l}{$\#$ Source's real magnitude = Background magnitude - $mag_s$, where background magnitude is flux due to background in mag/arcsec${}^2$}
\end{tabular}

\label{priortable}
\end{center}
\end{table*}

\subsection{The Lens Samples}\label{lenssampleI}
In this subsection we summarize the pilot sample selected for this paper.
Out of the 252 initially selected galaxies (Table \ref{sample}), 48 have $\rm M_{\star} > 10^{11} M_{\bigodot}$. The projected stellar masses are calculated within a cylinder of $3^{\prime \prime}$ diameter (see \citealt{auger2010b}) to keep the comparison consistent with SLACS (see Figure~\ref{flowchart}). From the remaining galaxies having $\rm M_{\star} < 10^{11} M_{\bigodot}$, we randomly select 11 galaxies motivated to test the performance of the pipeline. We perform lens-modeling on these two sets of samples. Given the pilot nature of the sample when comparing properties (e.g., total density slope) with observations we restrict to galaxies having $\rm M_{\star} > 10^{11} M_{\bigodot}$, also most reliable and least affected by SPH smoothing. To limit computation effort, we also currently only use one of the projected mass maps. The selected lenses cover nearly one dex in stellar mass  of the SLACS, but because of the limited volume of the simulations, they are poorly represented when approaching very massive ETGs. Finally we apply the end-to-end pipeline on the sample and analyze the results in this work. The result is that 34 out of 48 lenses having substantial arcs or Einstein rings (see Figure \ref{slacsvseagle}), converged to optimized solutions. 14 lenses having smaller arcs and more complex structure failed to converge to any reasonable solution in lens-modeling. Table~\ref{sample} summarizes the sample selection. 

The reason for our current down selection of the total sample is mainly due to the complexity in the implementation of automated lens-modeling with {\tt LENSED}. All the resulting mass maps, inverse magnification maps, convergence maps, the simulated lenses and model-fitting results are stored in a MySQL\footnote{\url{http://www.mysql.com/products/community/}} database, which has been widely used in astronomy (\citealt{lemson2006}). 



\begin{table*}
\caption{\normalsize The sample of EAGLE lenses used.}
\label{sample}
\begin{center}
\begin{tabular}{l l l l}
\hline
\hline
\multicolumn{4}{c}{\bf SEAGLE-I lenses}\\
\hline
Tag &No. of Galaxies & Proj. galaxies & \multicolumn{1}{c}{Comments}\\
\hline
A &252 & 756 & Total number of galaxies satisfying all the selection criteria (excluding $\rm M_{\star} > 10^{11}M_{\bigodot}$) of Table \ref{properties}\\
B&48 & 144 & Total number of galaxies satisfying all the selection criteria and having $\mathrm{M_{\star}>10^{11}\;M_{\bigodot}}$\\
C&48 & 48 & Number of modeled galaxies having $\mathrm{M_{\star}>10^{11}\;M_{\bigodot}}$ using one orientation\\
D&11 & 11 & Number of modeled galaxies having $\mathrm{M_{\star}<10^{11}\;M_{\bigodot}}$ for test purposes \\
\hline
\end{tabular}
\label{sample}
\end{center}
\end{table*}

\section{Lens-System Modeling}\label{lensmod}

Once we have created all the inputs to simulate mock lens systems including observational effect and masks, we model each lens system with {\tt LENSED} (\citealt{tessore15a}) using either an Elliptic Power Law (EPL; \citealt{tessore15b}) or a Singular Isothermal Ellipsoid (SIE; \citealt{kormann1994}) mass model, including external shear. A total of 14 and 15 parameters are sampled for the SIE and EPL models, respectively, and posterior distributions of all lens and source parameters are created via the MCMC method Nested Sampling. 

\subsection{Mass Models}

Various observational studies find that the EPL mass model (including the SIE) in general provides  a good approximation of the mass model of massive galaxy-scale strong gravitational lenses (\citealt{koopmans2006,koopmans2009,treu2004,newton2011}). As a first step we therefore model the lenses as a SIE plus external shear with the prior settings tabulated in Table~\ref{priortable}. The dimensionless surface mass density of the SIE model is given by 
\begin{equation}\label{siem}
\kappa(R)=\frac{b}{2R},
\end{equation} where $b$ is approximately the Einstein ring radius and $R$ is the elliptical radius defined by
\begin{equation}
R=\sqrt{q x^2 + y^2/q},
\end{equation}
where $q$ is the axis ratio (short over long axis length)and $x$, $y$ are cartesian coordinates of the model.
Similarly we model and analyze the lenses with an EPL mass model plus external shear, whose convergence is given by 
\begin{equation}\label{eqepl}
\kappa(R)=\frac{(2-t_L)}{2}\left( \frac{b}{R}\right)^{t_L} \qquad ,
\end{equation} 
where 0 < $t_L$ < 2 is the the power-law density slope of the mass model and the other parameters are the same as for the SIE model. This profile can arise from a three-dimensional mass distribution, given by
\begin{equation}\label{tslope}
\rho(r) \propto r^{-t}
\end{equation}
where $t=t_L+1$.

The EPL model allows us to (statistically) compare the ensemble of density slopes of the simulated lenses with those from SLACS and SL2S. We note that many of the SLACS density slopes were obtained from a combined lensing and dynamics analysis, not just from lensing. The same model also allows for a comparison with the convergence model fitting in Section~\ref{NMead}.

\subsection{Nested Sampling and Priors} 

We compare our models to the simulations using a Bayesian approach and sample the posterior via Nested Sampling (NS; \citealt{skilling2006,feroz2008,feroz2013}). NS is a modified Markov Chain Monte Carlo (MCMC) method that carries out the integral over the posterior probability distribution function (PDF) resulting in a value of the marginalized posterior, i.e. the evidence. As a by-product it also provides a first-order sampling of the posterior. The posteriors are used to estimate the maximum a posteriori (MAP) lens parameter values, their uncertainties, as well as potential degeneracies (see \citealt{tessore15a} for more details). The lens modeling is performed semi-automatically with 200 live points, where the initial values of priors are kept such that effectively all of the lens and source parameter spaces are covered (see Table \ref{priortable}). The parameter space is sometimes degenerate with multiple extrema, making a straightforward sampling difficult. To avoid any catastrophic failures in the reconstructed source or lens parameters, the analysis of the mock lenses was performed by trying a range of well-motivated priors without affecting the end result too much. A combination of rather uninformative Gaussian and uniform priors was found to be optimal in our modeling analysis. The details of the prior settings can be found in Table \ref{priortable}. In the EPL case we used tighter priors on $n_s$ and $q_s$ to avoid degeneracies that would slow the convergence. All NS chains are analyzed through {\tt GetDist}\footnote{\url{http://getdist.readthedocs.io/en/latest/}}. We get posterior distributions, corner plots and also marginalized plots for each individual source and lens parameters. We tested for a range of priors for the density slopes and shear (the two main parameters of the analysis) and found that our choice of priors improves convergence and reduces the computation time.  Hence we use the prior in Table \ref{priortable} to speed up convergence, but they have little to no impact on the final solution.  See Appendix \ref{Sapp} for details.

\subsection{Choice of Source Model}

The source parameters are observationally motivated (\citealt{bolton2008, newton2011})  and can in principle  be varied between sources. Our main goal in this paper is to infer global properties of the lensing galaxies, and the precise choice of the source model is currently of secondary importance (see Section \ref{lenscre}). So even-though some of the SLACS and SL2S sources show irregular morphologies, we expect a change in the source model not to bias the result especially since the systematic errors far outweigh random errors. \cite{tessore15a} for example performed rigorous testing to demonstrate that the choice of source model does not bias the lens-modeling (see Section 4.4 of \citealt{tessore15a}). They reported only a minor variance for the lens modeling parameters. We tested with a sub-sample of our pilot lenses and note sometimes an increase in the computational time for some but no change in the distribution of the parameters (see Appendix \ref{Sapp}). Hence in the current paper we have decided not to change the source model parameters between lenses.




\subsection{Convergence-Map Modeling}\label{NMead}

We also fit the EPL model in Equation~\ref{eqepl} directly to the convergence map of the galaxy inside the same mask that was used for the lens modeling, using the Nelder-Mead (NM) simplex method (\citealt{nm}) including some annealing to help convergence. We do this in order to compare the resulting lens-model parameters, as discussed in Section \ref{lensmod}, with those from the actual mass model of the simulated galaxy. Even the resulting parameters from this `direct' fit are still a limited representation of the true mass distribution, which can be more complex than an EPL mass model. Comparing the two, however, allows us to assess the reliability of the lensing results and the variance between the two parameter sets.  

We use an unweighted least square penalty function. We take the parameter values that minimize the penalty function from a set of ten different optimization runs with random initial parameter values, each having a maximum of 150 iterations. Most solutions agree well, with some outliers due to local minima. We choose the solution with the lowest penalty from this set, in general leading to a robust solution. This step of analysis in the pipeline is important for a number of reasons: (1) we obtain a fairly robust estimate of the main observables of the lensing galaxies such as the Einstein radius, axis ratio, position angle, and density slope, (2) we can make a direct comparison with the modeled output for each individual lenses, (3) the residuals obtained via this analysis could also be used for power-spectrum analysis.


\section{Comparing the Results from Lens and Convergence Mass Models}\label{results}
  
The two independent mass-model analyses i.e. via lens modeling and via direct convergence-map fitting, provide a consistency check and assessment of systematic errors on the simulated lenses when compared with observations (\citealt{marshall2007}).  
We compare the results from both model fits using an identical family of mass models. To compare their ensemble properties, however, we need to introduce a weighting scheme that mimics the selection effects in observed samples. These selection biases can be rather complex \citep{dobler2008} but in this paper we use the lens cross-section based on stellar mass. We ignore the magnification bias which is expected to change slowly with stellar mass for the most massive lens galaxies that we study. 

Below we discuss the results from the comparison between the two sets of parameters, in particular the complex ellipticity, its correlation with external shear, and the Einstein radius. We have used the SIE model results when comparing with observations' ellipticity and position angle which is consistent with the model used in SLACS and SL2S. For comparison of density slopes we have used EPL modeling results which are also consistent with the mass density slope model used in SLACS and SL2S.

\subsection{Complex Ellipticity}\label{2dcomplex}

\begin{figure*}
\includegraphics[width=0.99\columnwidth]{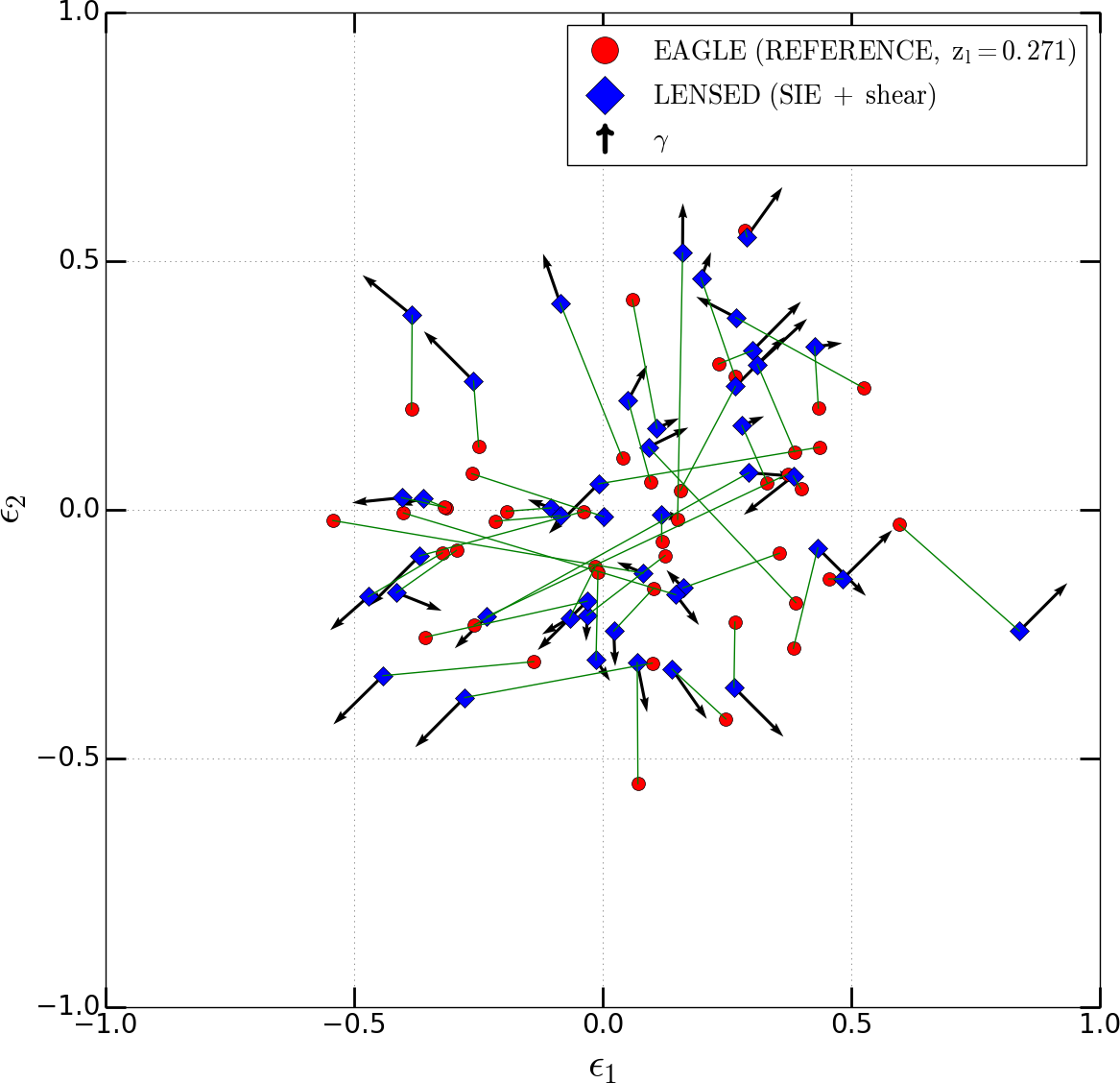}\hspace{0.2cm}
\includegraphics[width=\columnwidth]{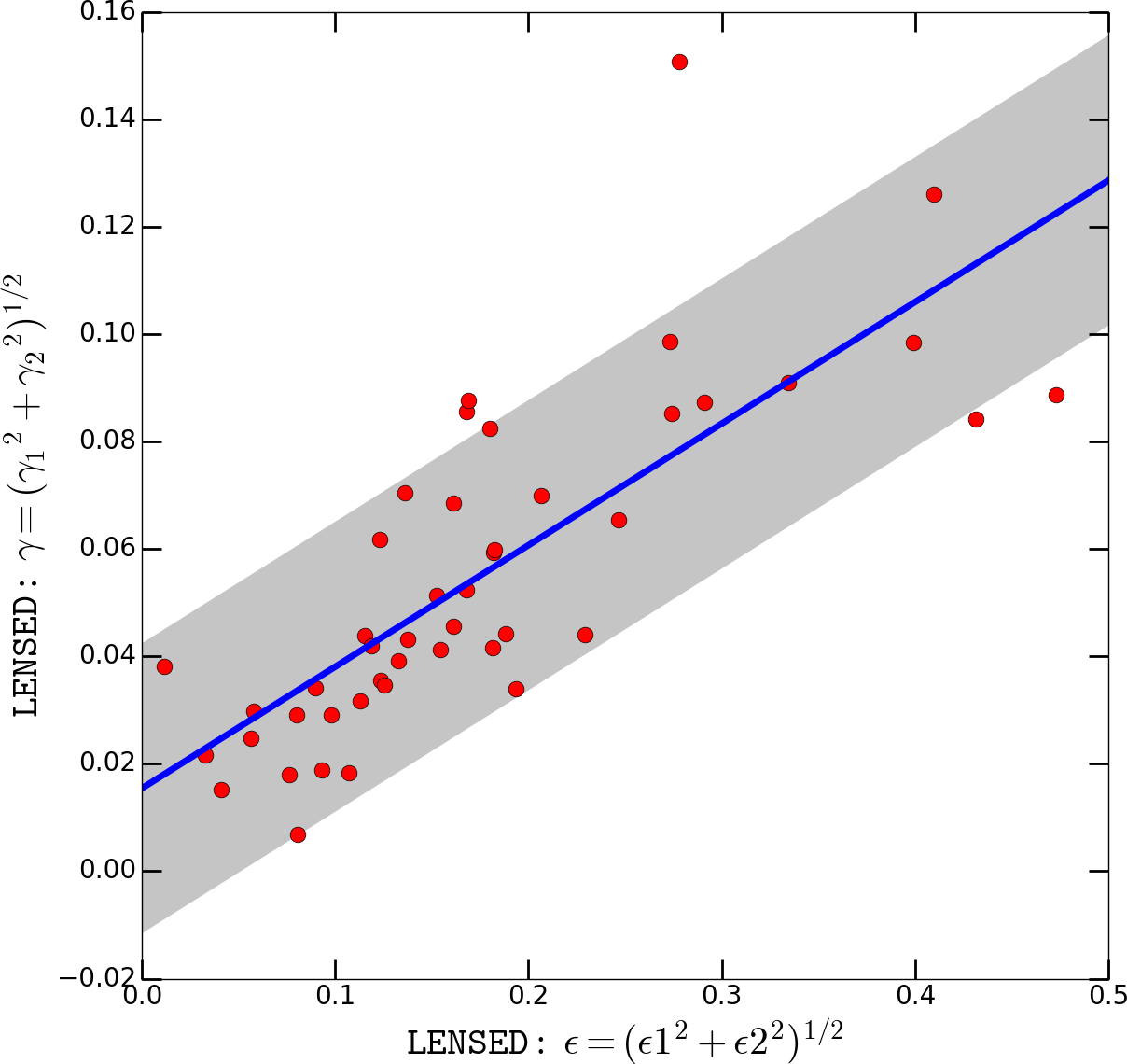}
\caption{\normalsize Left panel: The complex ellipticity (see eq.~\ref{es2}) of the SIE lens models (eq.~\ref{siem}) from {\tt LENSED} (blue diamonds) and from a direct fit to the convergence mass maps (red filled circles). The green line represents the line joining the two $\epsilon$ measurement ($\overline{\epsilon_{\rm L}\epsilon_{\rm K}}$), from lens-modeling and direct fitting. The shear vector ($\gamma$) tends to point radially outwards in this plot, so the ellipticity is degenerate with the shear. It is most likely to cause differences in the ellipticity in the direction of the shear which causes the true lens mass model to deviate from the assumed  mass models. Right panel: Complex ellipticity versus shear suggests a strong correlation among them. The shaded region shows the 1$\sigma$ (=0.027) interval. Here samples C and D (see Table \ref{sample}) have been used.}
\label{complexspace}
\end{figure*}

The position angle of the lens mass model has an ambiguity of $\pm \pi$ due to its point symmetry. In addition, when the lens is nearly round ($q \rightarrow 1$), the position angle becomes ill-defined. In order to disentangle this degeneracy we use a complex ellipticity representation which connects both $\phi$ and $q$.   

To accomplish this, we use the complex ellipticity defined as:
\begin{equation} \label{es1}
\epsilon ={ \frac{(1-q)}{(1+q)} e^{+2 i \phi }}\;,
\end{equation}
or in vector notation:
\begin{equation} \label{es2}
(\epsilon_1, \epsilon_2)^{\rm T} ={ \frac{(1-q)}{(1+q)} (\cos(2 \phi), \sin(2\phi))^{\rm T}}\;.
\end{equation}

In this representation rounder lenses will have a smaller values of $\epsilon$, regardless of the value of $\phi$. For smaller values of $q$, the absolute value of $\epsilon$ increases and $\phi$ should be better determined. The agreement between two models depends on the distance in this $\epsilon$-space, $\vert \epsilon_{\rm model\; 1}-\epsilon_{\rm model\; 2}\vert$. 

We present the model values of $\epsilon$ from both the lensing and direct fitting to the convergence maps in Figure \ref{complexspace}. The shear vectors from the lens model are also indicated. Calculating $\Delta \epsilon_1$ = $\epsilon_{\rm 1, L} - \epsilon_{1, \kappa}$ and  $\Delta \epsilon_2$= $\epsilon_{\rm 2, L} - \epsilon_{2, \kappa}$, where `L' suffix refers to {\tt LENSED} results and `$\kappa$' suffix refers to results from convergence map fitting, we find standard deviations of 0.24 and 0.17 for $\Delta \epsilon_{\rm 1}$ and $\Delta \epsilon_{\rm 2}$ respectively. The errors on the standard deviation are 0.1 and 0.07, respectively, hence the differences in the two directions are not significant. The scatter is significant though. 
We conclude that for lower stellar mass lenses ($\rm M_{\star} < 10^{11} M_{\bigodot}$) a significant difference can exist between the inference of the complex ellipticity from the convergence map and that from lens modeling. This might be regarded as a systematic error or bias in lens modeling which is hard to overcome. Below we investigate its cause in a little more detail. 

\subsection{Shear versus Ellipticity}

For the majority of the lens systems there is good agreement between the values of the complex ellipticity from both analyses (Figure~\ref{complexspace}, for errors see subsection~\ref{2dcomplex}), but some systems suffer from a significant mismatch. Some previous studies have associated the differences in alignment and ellipticity to the presence of external shear (complex $\gamma$), given by:
\begin{equation}\label{gammaeq}
\gamma=\gamma_1 + {\rm i}\gamma_2\; .
\end{equation}

\begin{figure}
\includegraphics[width=\columnwidth]{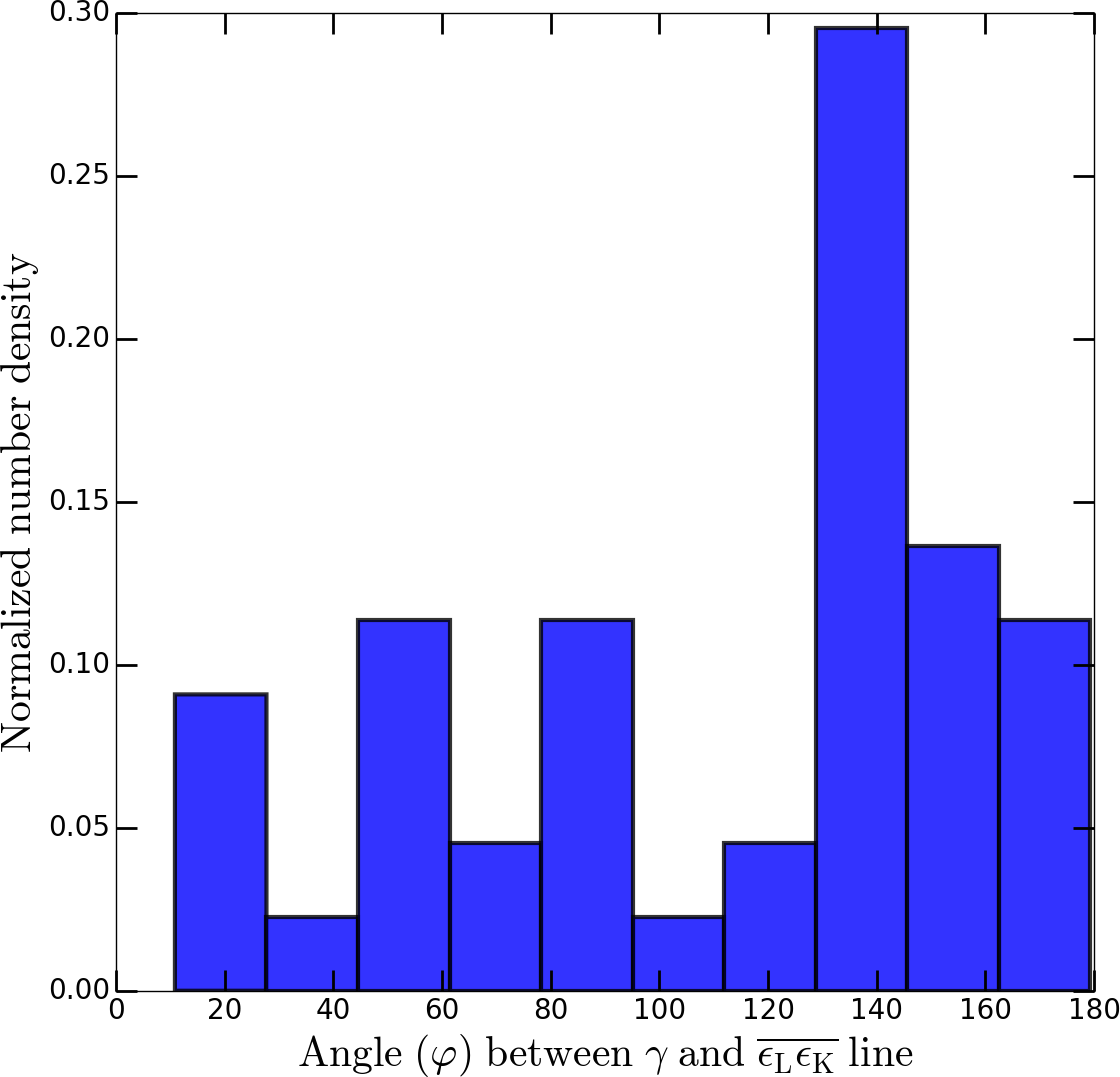}
\caption{\normalsize The normalized number density histogram of the angle ($\varphi$) between the shear vector ($\gamma$) and the $\overline{\epsilon_{\rm L}\epsilon_{\rm K}}$ line.}
\label{angle}
\end{figure}

They also indicated a pronounced degeneracy between ellipticity and external shear (\citealt{bandara2013, tessore15a}). We find that the  majority of the systems with large differences in the complex ellipticity between the lens and convergence modeling have external shears that have a preferred angle (Figure~\ref{angle}) to the vector joining the two $\epsilon$ measurement, $\overline{\epsilon_{\rm L}\epsilon_{\rm K}}$. This correlation in ellipticity and shear angles suggests that the `external' shear is in fact `internal' and is possibly caused by the mass distribution of the lens galaxy and not by external galaxies. In the latter case no strong correlation between shear and ellipticity angles would be expected. Hence, contrary to \citet{bandara2013}, who suggested that there may not be a direct correlation between $q$ and $\gamma$, we find a correlation  between $\gamma$ and $\epsilon$ values of our simulated lenses (Figure~\ref{complexspace}), being:
\begin{equation}
\gamma =0.226\epsilon+0.015
\end{equation}

Also we compute the angle ($\varphi$) between the shear vector and the line joining the complex ellipticities, $\epsilon_{\rm L}$ and $\epsilon_{\rm K}$, obtained from lens modeling and convergence ellipticity fitting respectively. Figure \ref{angle} illustrates the normalized distribution of angle $\varphi$ in degrees. It reaches peak at $\sim 135 \deg$ implying that the shear components $\gamma_{1}$ and $\gamma_{2}$ appear orthogonal to $\epsilon_{1}$ and $\epsilon_{2}$ respectively. But the standard deviation in the distribution is $\sim$ 50 $\deg$ which also suggests that orthogonal orientations of the shear vector have considerable scatter.


Based on this strong correlation and the apparent alignment or orthogonality between shear and ellipticity, we conclude that much of the difference in ellipticity inferred from lens models and direct fitting to the convergence maps is the result of an internal shear causing a bias in the lens models. This shear is therefore not caused by the external galaxies, but more likely by a difference between the assumed mass model (SIE) and the true mass model. Its difference is likely compensated for by the shear used in lens modeling. A first order deviation could be boxy or diskines of the galaxy.

\subsection{Einstein Radius}
\begin{figure}
\includegraphics[width=\columnwidth]{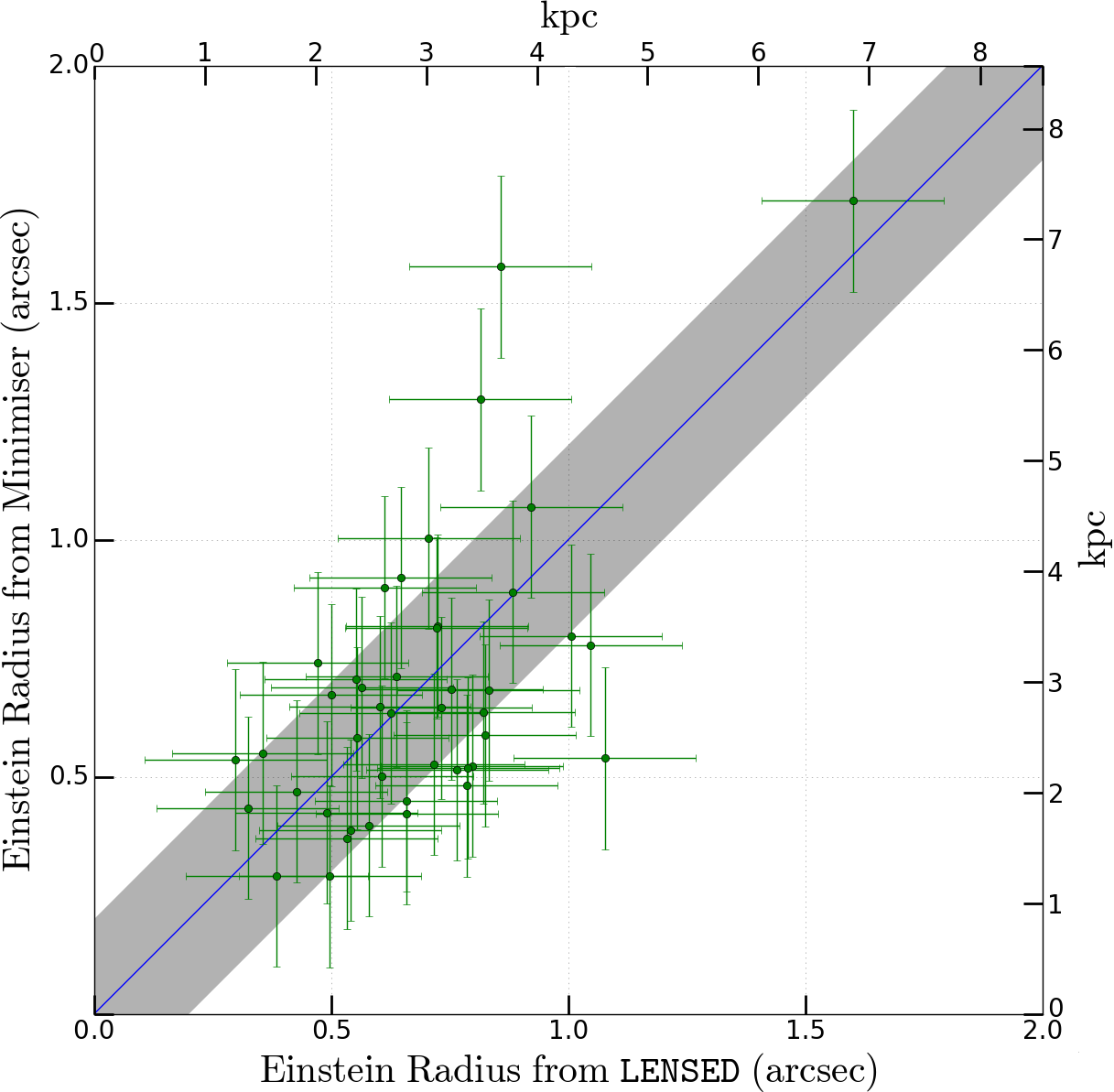}
\caption{\normalsize Comparison between the values of the Einstein radius $\rm R_{Eins}$ inferred from SIE lens modeling and convergence fitting. The blue line is the one-to-one correspondence. The scatter is given by the gray shaded region. The error-bars are the same 0.2 value of the scatter. Here samples C and D (see Table \ref{sample}) have been used.}
\label{eradius}
\end{figure}

Another comparison between the two models is that between the inferred Einstein radii. We have to be careful here though since the lens model that we use is a singular mass model whereas the convergence is affected by the SPH kernel and therefore has a small (0.7\,kpc) core that might affect a direct comparison. Figure \ref{eradius} shows the comparison of the Einstein radii obtained from the convergence and the lens modeling. The values obtained from the two independent analyses agree reasonably well and without an appreciable bias, but there is a large $\sim$20\% scatter (shaded region) from the one-to-one line. We rejected four data points which have a difference of more than $0.5''$ as critical failures that can heavily bias the standard deviation. From individual inspection of the first, we find that the  Einstein radii from the lens models seem more reliable than from the convergence fitting, possibly due to the central core affecting a direct fit.

\subsection{Density Profile}

\begin{figure}
\includegraphics[width=\columnwidth]{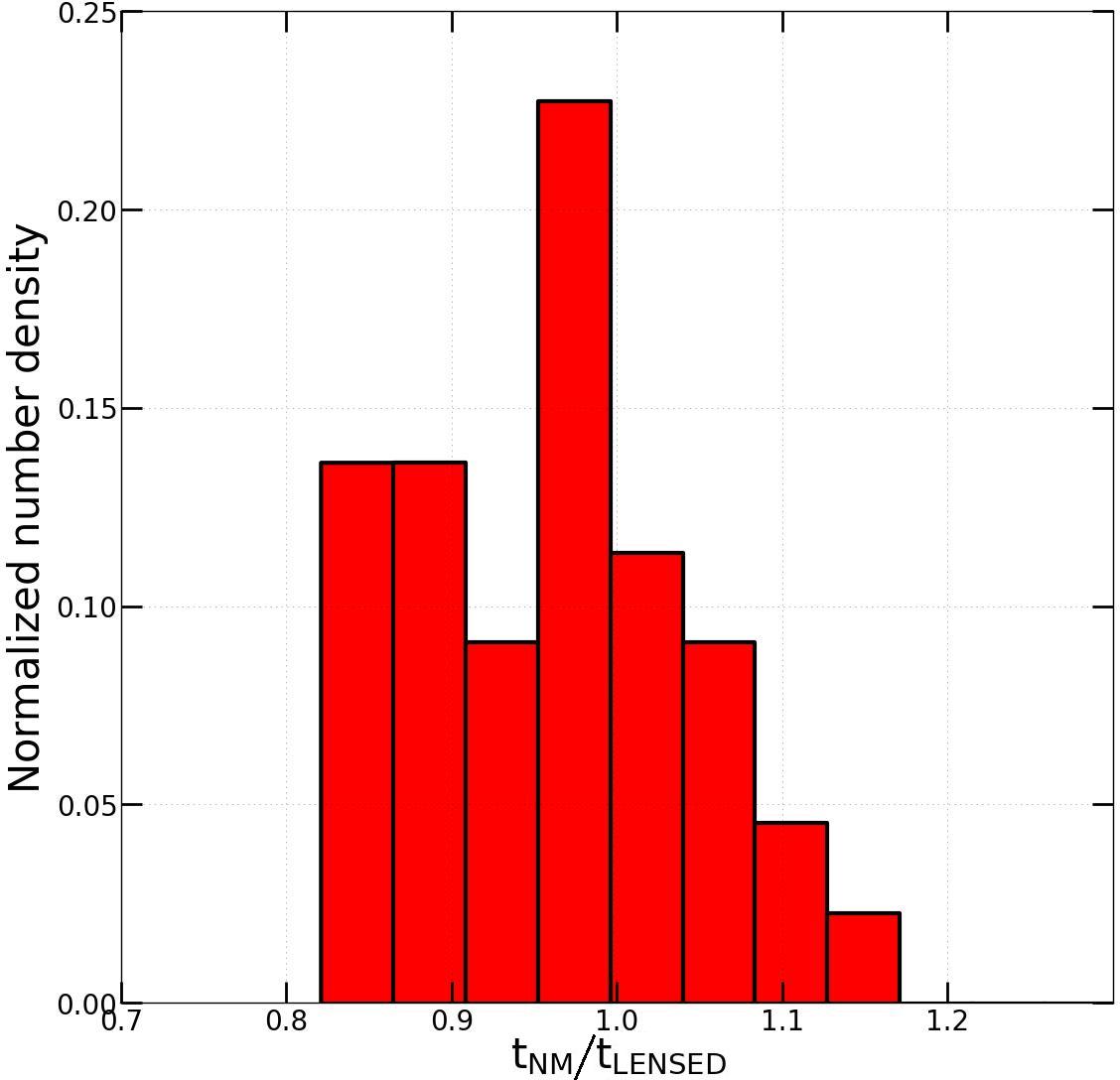}
\caption{\normalsize Comparison between the values of the mass density slope obtained from {\tt LENSED}, $\rm t_{LENSED}$, and convergence fitting, $\rm t_{NM}$. Here samples C and D (see Table \ref{sample}) have been used.}
\label{tdtl}
\end{figure}
Finally we describe the comparison of surface density slopes inferred via convergence fitting, $\rm t_{NM}$ and {\tt LENSED}, $\rm t_{LENSED}$ respectively. In Figure \ref{tdtl}, we show a normalized number density histogram of ratio of the mass density slopes analyzed from both the processes.
We find a mean ratio of 0.91 for $\rm t_{NM}$/$\rm t_{LENSED}$, with a standard deviation of 0.17. Eventhough the mean suggest a one-to-one correlation between the mass density slopes obtained from lens modeling and convergence fitting, from Figure \ref{tdtl} we can also see a tail suggesting that some differences are still present in them. These differences can be attributed to the different methodologies used in direct fitting and lens modeling. The lens modeling fits the density profile (more precisely that of the potential) near the lensed images, whereas the direct fit is mostly fitting the higher density regions inside the mask. The overall agreement however is encouraging, suggesting that lensing does not provide strongly biased density slopes. 

\section{Comparisons with SLACS and SL2S}\label{comparisons}

\begin{figure}
\begin{center}
\includegraphics[width=\columnwidth]{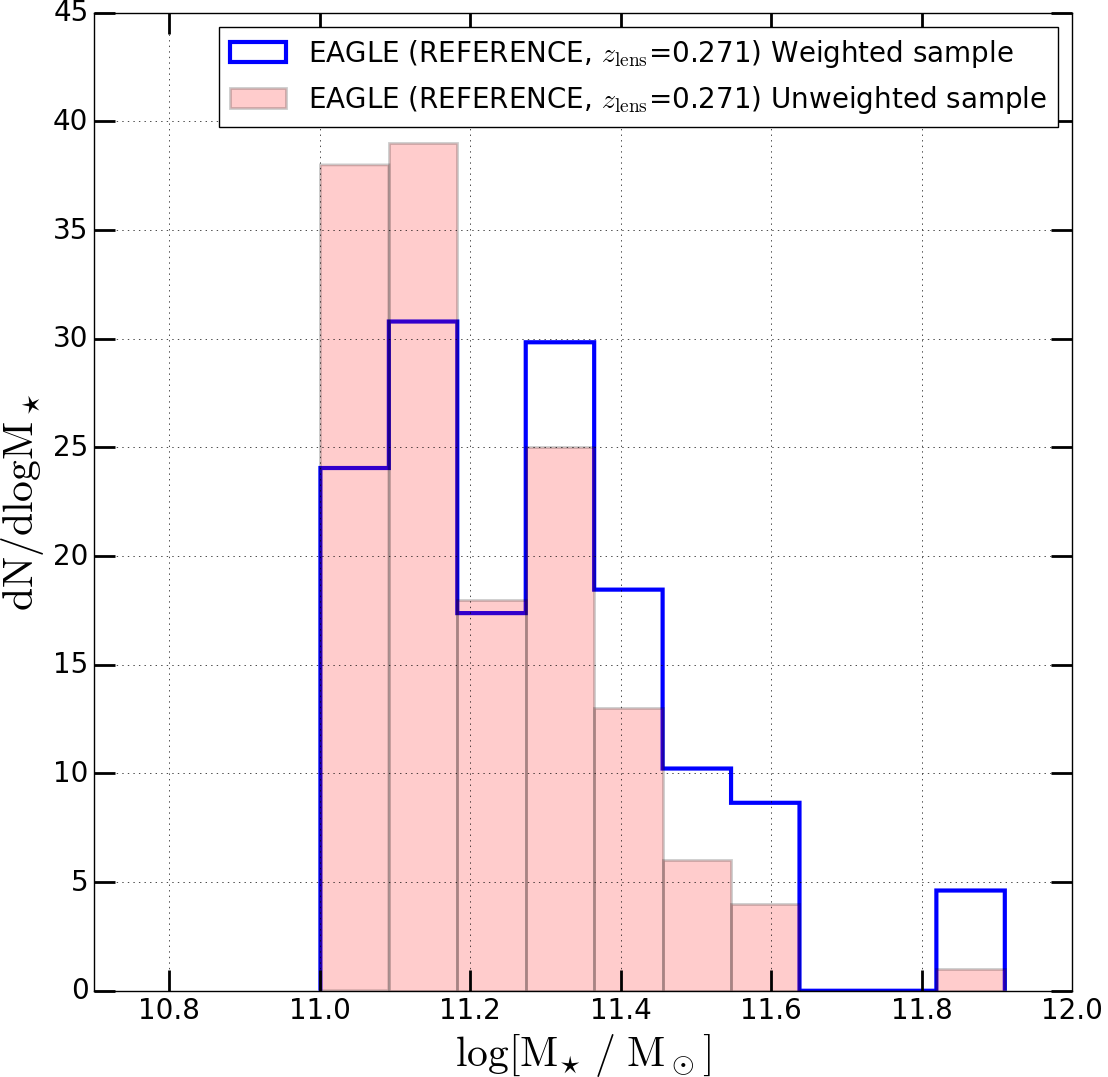}
\end{center}
\caption{\normalsize The mass function of galaxies having stellar masses $\rm{M_\star >10^{11}M_{\bigodot}}$, including and excluding the weighting scheme based on stellar mass as discussed in the 
text. Here sample B has been used.}
\label{weight}
\end{figure}

Having studied how well lens-model parameters agree with direct fitting of the same surface density model to the simulations, and having assessed their level of systematic and/or random differences, in this section we do a first-order comparison between EAGLE lenses from the Reference model with those from SLACS and SL2S. In the latter cases we make a correction for the lensing cross-section inferred from their stellar masses (see Section \ref{weightscheme}). We concentrate on lenses with a stellar mass exceeding $10^{11}$\,M${}_{\odot}$, which we believe are currently most reliably represented in the EAGLE simulations, based on the assessments in the previous section.  

\subsection{SLACS \& SL2S}

SLACS is a HST snapshot imaging survey, where lens candidates were selected spectroscopically from SDSS \citep{bolton2006}. With more than a hundred confirmed strong lens systems, SLACS is currently the largest and most complete early-type lens survey. The SLACS candidates were selected to yield bright lenses i.e. massive ETGs, in particular Luminous Red Galaxies (LRGs) with faint star-forming background sources, generally with irregular morphology. Hence the SLACS sample was primarily a lens-selected sample. The approximate mean Einstein radius is 1.2 arcsec \citep{koopmans2006,auger2010a} with background galaxies having a typical scale length of about 0.2 arcsec \citep{koopmans2006}. In later SLACS papers the sources were modeled with S\'{e}rsic profiles \citep{newton2011}. 

SL2S (\citealt{cabanac2007}) is a survey dedicated to find and study galaxy-scale and group-scale strong gravitational lenses in the Canada France Hawaii Telescope Legacy Survey (CFHTLS). The galaxy-scale SL2S lenses are found by searching the 170 square degrees of the CFHTLS with the automated software {\tt RingFinder} (\citealt{gavazzi2014R}) looking for tangentially elongated blue arcs and rings around red galaxies. The lens candidates undergo a visual inspection and the most promising systems are followed up with HST and spectroscopy. For details one can consult \cite{gavazzi2012}.

SL2S differs from SLACS in the way lenses are found. While in SL2S lenses are identified in wide-field imaging data, SLACS lenses were selected by searching for spectroscopic signatures coming from two objects at different redshifts in the same line of sight in the Sloan Digital Sky Survey (SDSS) spectra. These two different techniques lead to differences in the population of lenses in the respective samples. Due to the relatively small fiber used in SDSS spectroscopic observations ($1.5^{\prime \prime}$ in radius), the SLACS spectroscopic survey tends to limit the search to lenses with equivalent or smaller Einstein radii, where light of both the arcs from the lensed source and the deflector are captured within the fiber. SL2S however finds a larger number of lenses with Einstein radii greater than $1^{\prime \prime}$, because  they are more clearly resolved in ground-based images. BELLS have used the same methodology as SLACS to select the strong lenses, so they do not provide additional information on selection effect, hence they are not included in this comparison (\citealt{Brownstein2012}).

Figure \ref{slacsvseagle} shows a subset of simulated lens systems closely mimicking SLACS lenses \citep{bolton2006} in morphology and largely being arc and ring systems. Small-scale structure in the lensed images is lacking, because we are using a S\'{e}rsic source rather than the more complex (star-forming) real systems. We do not aim to reproduce small-scale features in the source because we only compare global properties such as Einstein radii, axis ratios, density slopes and position angles between SLACS and SL2S and the recovery of these quantities should not strongly depend on the fine-scale structure of the source. 

\subsection{Lens Selection Bias}\label{weightscheme}

The statistical comparison of a sample of volume and mass-selected lenses systems from simulations with observations is difficult due to selection biases as well as the often small simulation volumes compared to the volumes probed by lens surveys. The sample properties are for example affected by a lens cross-section that is mass dependent and a magnification bias which are different for different surveys. Because a precise analysis is beyond the scope of this paper, we only correct for the largest of these effects, being the lens cross-section. We assume that the magnification bias does not vary strongly with galaxy mass, which is a reasonable assumption if the source is small compared to the lens cross-section and if the properties of the lens mass model (besides mass) such as its flattening also do not depend strongly on mass. In most surveys that are dominated by $\rm M_\star$ early-type galaxies, these are reasonable assumptions. 

The lensing cross-section for the EPL model that we assume (generally close to the SIE), is proportional to the square of the Einstein radius, which in turn is proportional to the stellar mass, assuming the Faber-Jackson relation (\citealt{faber1976}) and a constant mass-to-light ratio. The latter is a direct observable in both the simulations and observations. 
We therefore define our weighting scheme (Figure \ref{weight}) per lens simply as
\begin{equation}\label{w2}
\centering
W(M)\equiv\left( \frac{M_\star}{\langle M_\star \rangle} \right),
\end{equation}
with $\langle M_\star \rangle$ being the average of the sample.
This scheme is used to re-weigh each strong lens when comparing distributions of parameters between observed lenses (i.e. SLACS and SL2S) and simulated lenses. Because most of the lenses are drawn from the exponential tail of the mass function, even such a rather strong (linear) reweighing has only a limited impact on the tilt of the distribution functions. Figure~\ref{weight} shows that although massive ellipticals are rarer than low stellar mass galaxies, more massive ellipticals are more likely to be observed in a lens-survey because of their larger lensing cross-section. 

Now we compare the properties of simulated EAGLE lenses with SLACS and SL2S. In this paper we restrict ourselves to $\rm M_\star > 10^{11}\; M_{\odot}$. Table \ref{sample} summarizes the number of galaxies, lenses and projected mass maps. We compare the number density versus stellar mass, the mass density slope and then compare ellipticity and position angle in complex space that we mentioned in section \ref{2dcomplex}.

\subsection{Stellar Masses}

\begin{figure*}
\begin{center}
\includegraphics[width=\columnwidth]{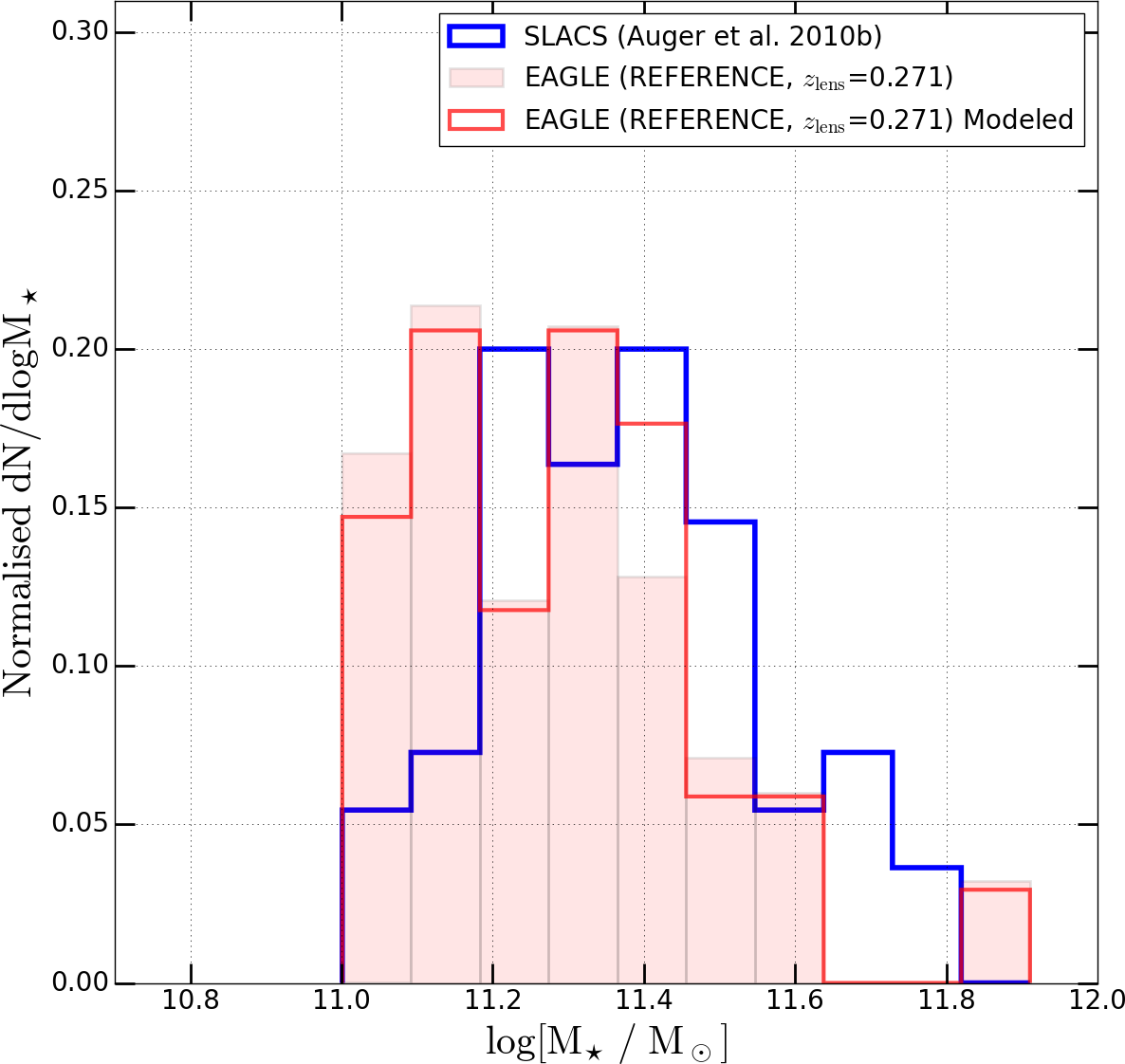}
\includegraphics[width=\columnwidth]{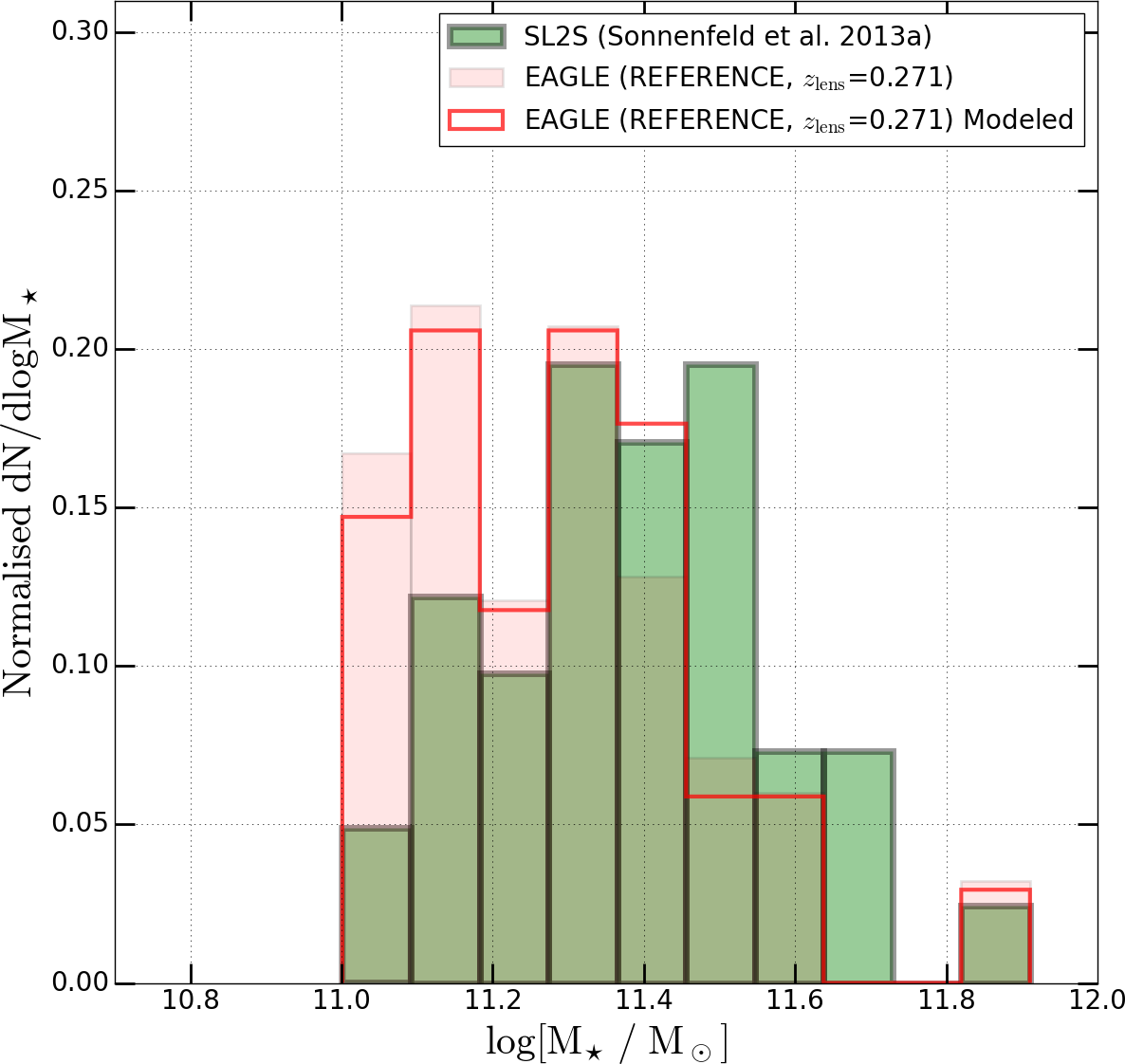}
\end{center}
\caption{\normalsize Comparison of the EAGLE lenses (total sample and modeled sample) with SLACS (left) and SL2S (right) lenses having stellar masses $\rm{M_\star >10^{11}\;M_{\bigodot}}$.}
\label{Mscomp}
\end{figure*}

In Figure \ref{Mscomp} we compare the simulated and observed stellar mass functions of the lens samples. Although not perfect, given the small-number statistics of the samples, the distributions show that the re-weighting scheme results in a distribution of EAGLE lenses similar to that of SLACS and SL2S. Although a significant number of EAGLE lenses are within the stellar mass range $\rm 10^{11.0-11.2}\;M_{\bigodot}$ which are not  common in SLACS and SL2S, we can still compare them considering that the simulation box covers only a fraction of the real universe and sample variance is thus very large.

\subsection{Density Slopes} 
\begin{figure*}
  \centering
  \subcaptionbox*{}[\columnwidth][c]{%
   \includegraphics[height=1.2\textwidth,width=0.9\columnwidth]{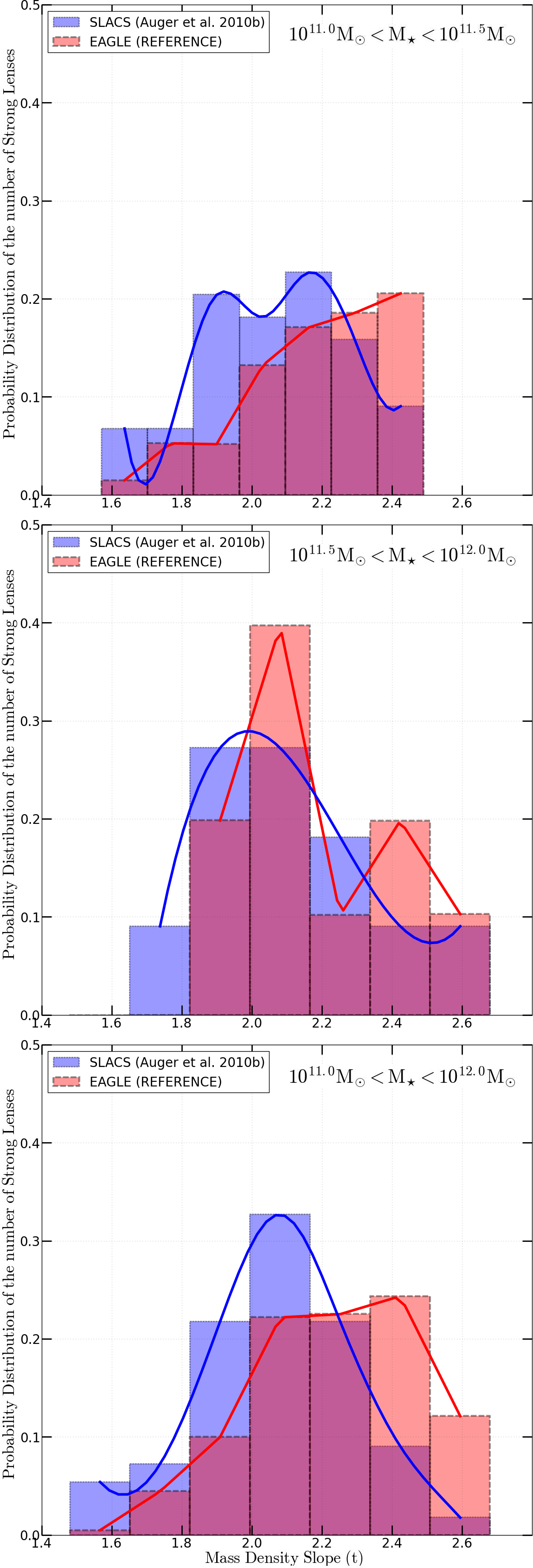}}
 \subcaptionbox*{}[\columnwidth][c]{%
    \includegraphics[height=1.2\textwidth,width=0.9\columnwidth]{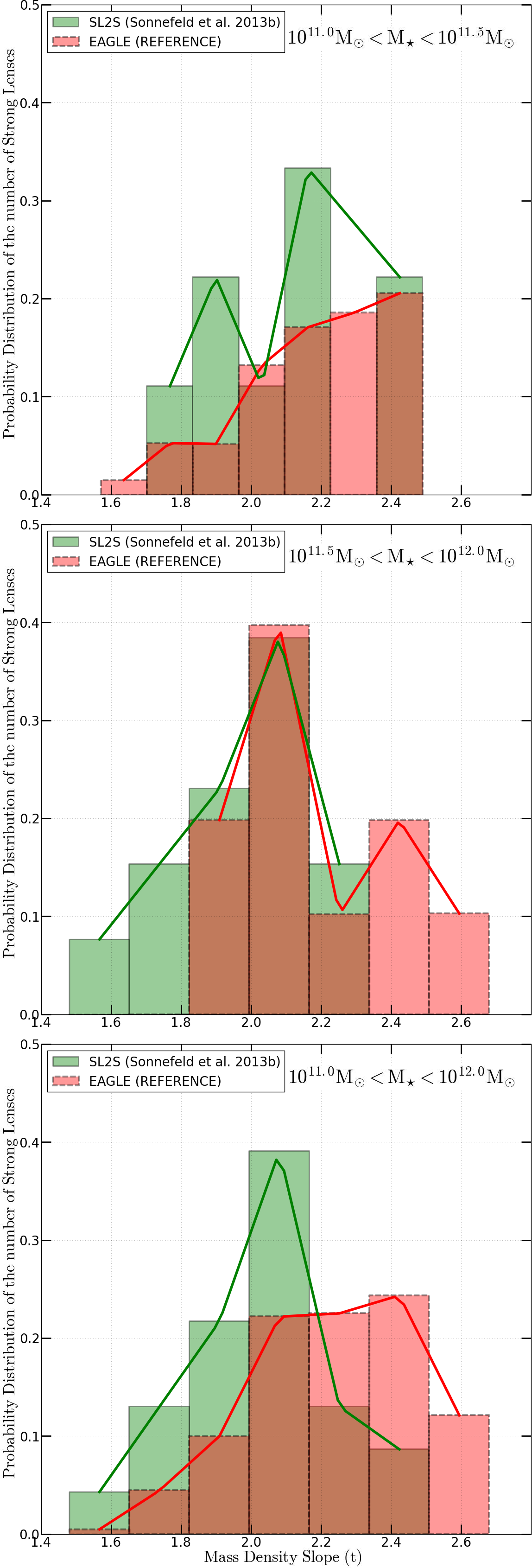}}
 
 \caption{\normalsize The probability distributions of the mass density slope ($t$, see eq. \ref{tslope}) for selected lenses of Reference scenario at $\rm z_l=0.271$ in three stellar mass-bins of $\rm 10^{11.0-11.5} \;M_{\bigodot}$,  $\rm 10^{11.5-12.0}\; M_{\bigodot}$ and $\rm 10^{11.0-12.0} \;M_{\bigodot}$. They are compared to SLACS (left column) and SL2S (right column) samples. The distributions for EAGLE lenses have been weighted using Equation \ref{w2}. We have created a homogeneous and statistically representative sample of simulated mock strong lens systems mimicking observational surveys of SLACS and SL2S.}
\label{comparisionslope}
\end{figure*}


To compare the density slopes, $t$ (see eq. \ref{tslope}) of the simulated lenses with their SLACS and SL2S counterparts, we have binned them into two mass ranges and one overlapping mass range: $\rm 10^{11.0-11.5} M_{\odot}$, $\rm 10^{11.5-12.0} M_{\odot}$ and $\rm 10^{11.0-12.0} M_{\odot}$. Figure \ref{comparisionslope} shows the (normalized) histograms of the density slope. We find a mean value of density slope of $2.26$, which quite similar, although slightly higher than SLACS with $2.08$ and SL2S with $2.16$. This can be explained by several SLACS lenses having much more shallower slopes ($\approx$1.6) especially in the $10^{11.5-12.0}$~M${}_{\odot}$ stellar mass range and SL2S lenses are highly concentrated around density slope $\approx$2.10 in all the three mass-bins which makes the mean value of SLACS and SL2S density slope lower than that of EAGLE. This slight difference (although within rms limits) can be attributed to the subgrid physics, feedback mechanisms used in this simulation run and/or due to systematics. These aspects will be tested in the forthcoming paper in this series. 

The mean slope is also consistent with other studies where the mass density slopes are determined from the central dynamics of local galaxies (\citealt{Dutton_Treu14,Tortora14}) and recent simulations  (\citealt{remus2017,xu2017}). In these simulations, however, the slopes were calculated directly from the particle distributions whereas here we use lens modeling and convergence fitting.
In Table 5 we have summarized the mean, root mean square (RMS), median and the 68\% confidence interval for the three stellar mass bins.

\subsection{Complex Ellipticity} 

\begin{figure*}
\begin{center}
\includegraphics[width=\columnwidth]{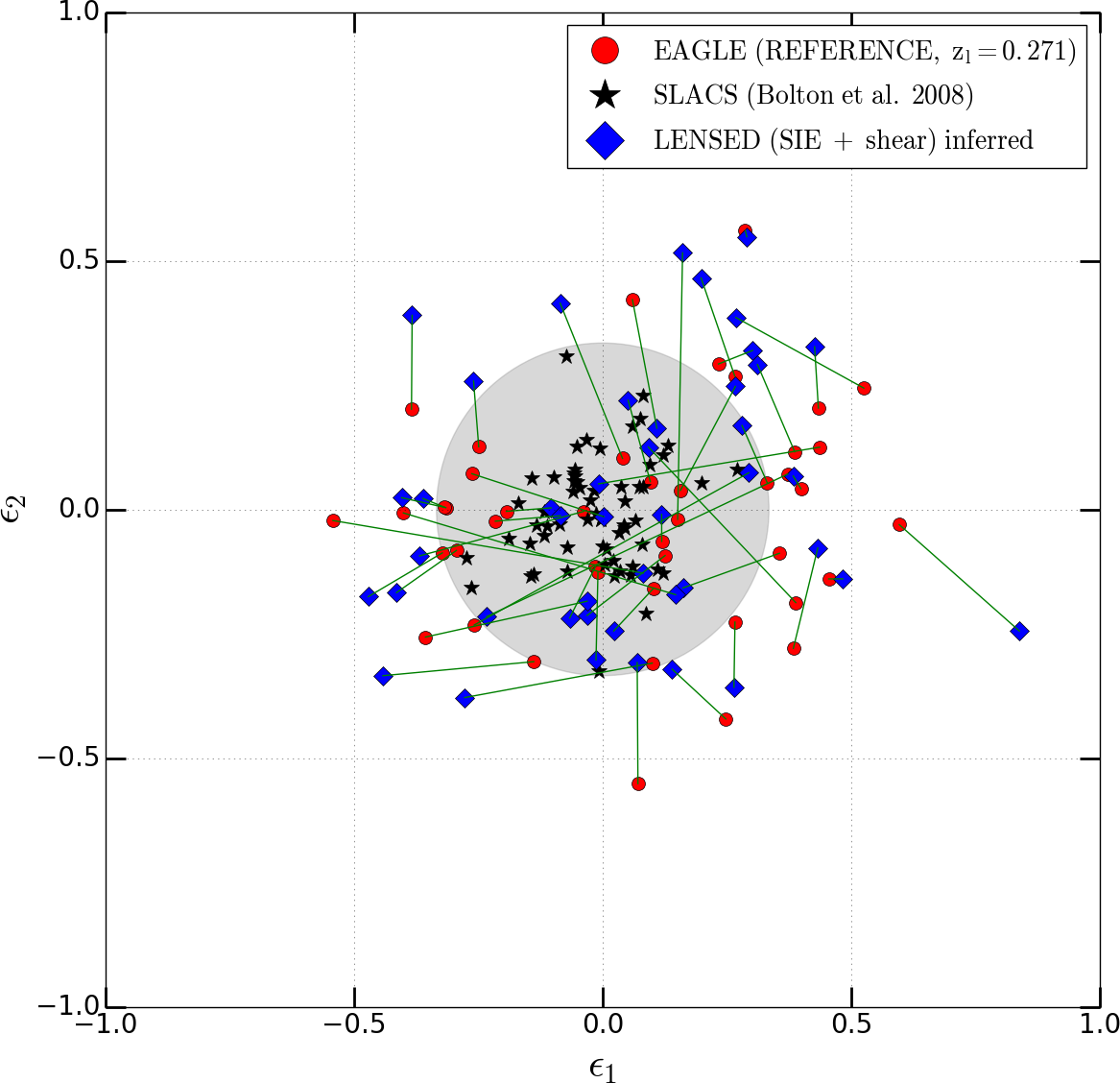}\hspace{0.2cm}
\includegraphics[width=\columnwidth]{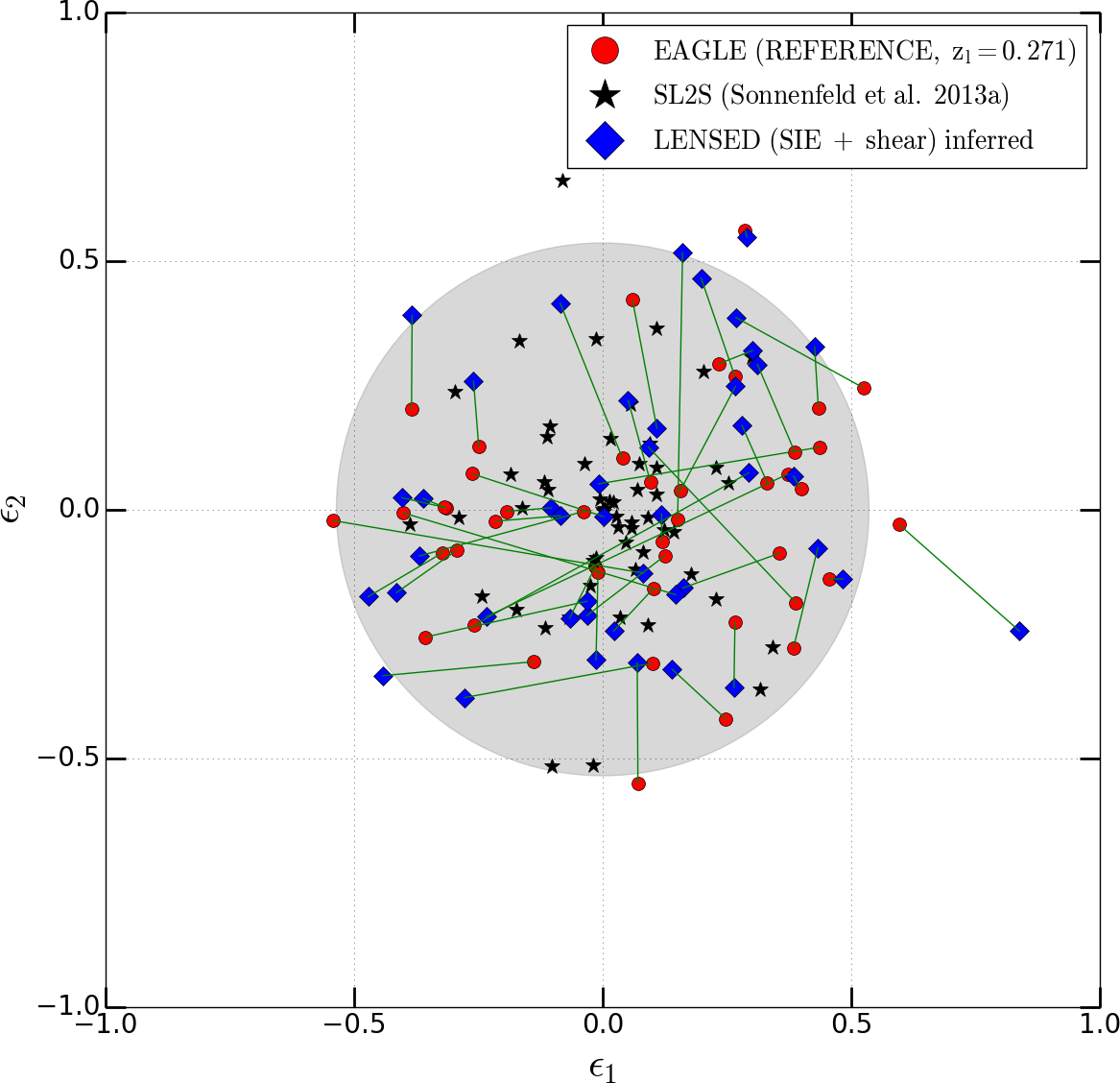}
\end{center}
\caption{\normalsize Comparison of the axis ratio and position angle of the EAGLE lenses (total modeled sample) with SLACS (left panel) and SL2S (right panel). The shaded region shows the maximum spread of the SLACS and SL2S lenses in left and right panels respectively. 12 out of 45 EAGLE lenses fall completely outside of the SLACS range. Whereas only 2 out of 45 EAGLE lenses are beyond SL2S range. This indicates that the modeled parameters for SL2S are more degenerate than SLACS or might also intrinsically be having a larger spread in ellipticity due to their group environment. We have rejected one SL2S data point (SL2SJ221326-000946) [located at the corner of the legend box in the right panel plot] which we consider to be a critical failure in SL2S comparison plot (see section 6.5).}
\label{Scomp}
\end{figure*}

Finally, we compare the complex ellipticity from lens modeling of EAGLE lenses (Section \ref{2dcomplex}) and SLACS. We do not use SL2S results since we do not have direct access to these mass model parameters. Figure \ref{Scomp} shows the SLACS lenses in black dots and EAGLE lenses in a similar way as in figure \ref{complexspace}. The gray shaded region shows the domain of SLACS obtained from \cite{bolton2008a}. We find broad agreement between them for 33 out of 45 modeled EAGLE lenses. SLACS lenses are concentrated around the origin of the plot but still some of them suffer from the $q-\phi$ degeneracy (or \say{conspiracy}), with 12 out of 45 modeled EAGLE lenses are completely outside of the shaded region (Figure \ref{Scomp} left panel). When comparing with SL2S lenses, we see a spread in complex ellipticity for SL2S data whereas most EAGLE lenses are well within the range. This might indicate that some of the SL2S lenses suffers from large modeling degeneracy. But also SL2S lenses reside in a group environment. So the environment of SL2S lenses can also contribute to the cause of having a much broad parameter space for axis-ratio and ellipticity for SL2S lenses. The most critical case being SL2SJ221326-000946 which is a disky system with axis ratio 0.20 and P.A. -41.5 $\deg$ (measured east of north).

\begin{table}
\begin{center}
\caption{\normalsize The mean, rms, median and 68\% confidence limits of mass density slopes, $t$ (see eq. \ref{tslope}) of the simulated lenses.}
\label{slopeprop}
\begin{tabular}{l l l l l }
\hline
\hline
$\rm \log M_\star \;(M_{\odot})$&Mean & RMS & Median& 68\% CL\\
\hline
$\mathrm{11.0-11.5}$&2.26&0.26&2.26&1.49-3.03\\
$\mathrm{11.5-12.0}$&2.28&0.21&2.23&1.46-3.00\\
$\mathrm{11.0-12.0}$&2.26&0.25&2.26&1.49-3.03\\
\hline
\end{tabular}
\end{center}
\label{slopeprop}
\end{table}


\section{Discussions and Summary}\label{discussions}

In this paper we have presented an end-to-end strong-lens simulation and modeling pipeline, allowing us to assess the (dis)agreement between mass-model parameters (e.g.\ density slope, complex ellipticity) inferred from lens modeling and from direct fitting to the simulations, using the same mass-model family. In the current implementation (called \say{SEAGLE}), we use the EAGLE (Reference-L050N0752) hydrodynamical galaxy simulations (\citealt{s15,c15}), the {\tt GLAMER} ray-tracing package (\citealt{metcalf2014,petkova2014}), the {\tt LENSED} lens-modeling code (\citealt{tessore15a}), and model all lenses as power-law elliptical mass models or singular isothermal ellipsoid mass models with external shear. 

When making a stellar mass cut in EAGLE at $> 10^{11} \rm M_{\odot}$ and after re-weighting the EAGLE stellar mass function $\rm dN/dM_{\star}$ by a simple estimator of the lens cross-section (Figure~\ref{weight}), we find that the simulated lenses have a broadly similar stellar mass function to SLACS and SL2S. Their visual appearance is also strikingly similar (see Figure \ref{slacsvseagle}). This motivates us to compare these observed lens samples to the simulated lens systems. 

In more detail, the conclusions from this work can be briefly summarized as follows:

(1) When comparing the results from lens modeling and direct fitting of the mass surface density of lenses in the simulations,  we find a correlation between the external shear ($\gamma$) and the complex ellipticity ($\epsilon$), with $\gamma \sim \epsilon/4$ (Figure~\ref{complexspace}). This correlation indicates a degeneracy in the mass model, where the shear compensates for a mismatch between the model and the real mass distribution. This is supported by the fact that the shear and complex ellipticity angles are correlated (Figure~\ref{angle}). This could be related to a disky or boxy mass model, ill described by the elliptical model in the direct fit, but affecting the lensed images. 

(2) The Einstein radii of the lens models and direct fits broadly match, i.e.\ within a 20\% scatter (Figure~\ref{eradius}). We attribute this surprisingly large scatter due to the fact that lens modeling really only fits the density profile (more precisely that of the potential) near the lensed images, whereas the direct fit is mostly fitting the higher density regions inside the mask, which might lead to a larger scatter when inferring the Einstein radius. We see no significant bias however and believe that the scatter is largely coming from the convergence fits. 

(3) From the EAGLE Reference model we find that the mass density slope of galaxies inferred from lens-modeling ($\rm t_{LENSED}$) and direct fitting ($\rm t_{NM}$) generally agree well with the ratio, $t_{NM}/t_{LENSED}$, having a mean of 0.91 and standard deviation of 0.17 (Figure~\ref{tdtl}). 

(4) The lens modeling yields a mean density slope of $t = 2.26$ (an SIE has $t=2$). Direct fitting, though, shows that this slope has a typical rms of $0.15$ with that from lensing, setting a limit to the level to which the density slope can be determined (at least in these simulations). The average total density slope is higher than for SL2S, $t=2.16$ or SLACS, $t=2.08$ (Figure~\ref{comparisionslope}). This slight difference within rms can be due to the feedback mechanisms and sub-grid physics adopted in simulations, and also due to systematics.

(5) The complex ellipticity of EAGLE and SLACS lenses shows that three quarters of the modeled EAGLE lenses agree quite well with the distribution of SLACS lenses which is shown by the shaded region (Figure~\ref{Scomp}). Ten out of twelve of the more elliptical simulated lenses have stellar mass $\rm < 10^{11}M_{\odot}$. Modeled complex-ellipticities for SL2S lenses are much more degenerate and EAGLE results are well within the SL2S lens domains. This larger ellipticity of SL2S lenses might also be due to the group environment in which the lenses resides. Although a degeneracy exists between $q$ and $\phi$ but for massive ETGs in EAGLE we find broad agreement with SLACS and SL2S lenses.


In this work we have presented a pipeline to create and model simulated realistic mock strong lenses and a pilot comparison between EAGLE lenses and SLACS and SL2S lenses. Even though previous work (e.g. \citealt{xu2012}) have simulated lenses and tested lensing degeneracies, we have extended those studies by incorporating the aspects of lens modeling and by comparing the inputs to quantify systematic effects in lens modeling. Moreover this work also aims at a full automation of simulated lens creation, modeling and comparison with observation which will be needed when future surveys start discovering 1000s of strong lenses. In the future, we will use the SEAGLE pipeline to analyze various galaxy formation scenarios of EAGLE, and compare them to observations in order to disentangle various aspects of galaxy formation mechanisms.





\section*{Acknowledgements}
We thank the anonymous referee for her/his insightful comments and constructive suggestions which improved this work to its present form.
SM, LVEK, CT and GV are supported through an NWO-VICI grant (project number 639.043.308). SM thanks SURFSARA network in Amsterdam. SM thanks Saikat Chatterjee, Dorota Bayer, Cristiana Spingola, Hannah Stacey and Carlo Enrico Petrillo for useful feedback on the representation  of the paper. NT acknowledges support from the European Research Council in the form of a Consolidator Grant with number 681431. MS is supported by VENI grant 639.041.749. JS is supported by VICI grant 639.043.409. RAC is a Royal Society University Research Fellow. FB thanks the support from the grants ASI n.I/023/12/0 ``Attivit\`a relative alla fase B2/C per la missione Euclid'' and PRIN MIUR 2015 ``Cosmology and Fundamental Physics: Illuminating the Dark Universe with Euclid''. This work used the DiRAC Data Centric system at Durham University, operated by the Institute for Computational Cosmology on behalf of the STFC DiRAC HPC Facility (www.dirac.ac.uk). This equipment was funded by BIS National E-infrastructure capital grant ST/K00042X/1, STFC capital grant ST/H008519/1, and STFC DiRAC Operations grant ST/K003267/1 and Durham University. DiRAC is part of the National E-Infrastructure.  This work was supported by the Netherlands Organisation for Scientific Research (NWO), through VICI grant 639.043.409.



\bibliographystyle{mnras}
\bibliography{./SEAGLE-I/references} 
\appendix \label{app}

\section{Cores in Simulations and masking}\label{Eapp}
\begin{figure}
\includegraphics[width=\columnwidth]{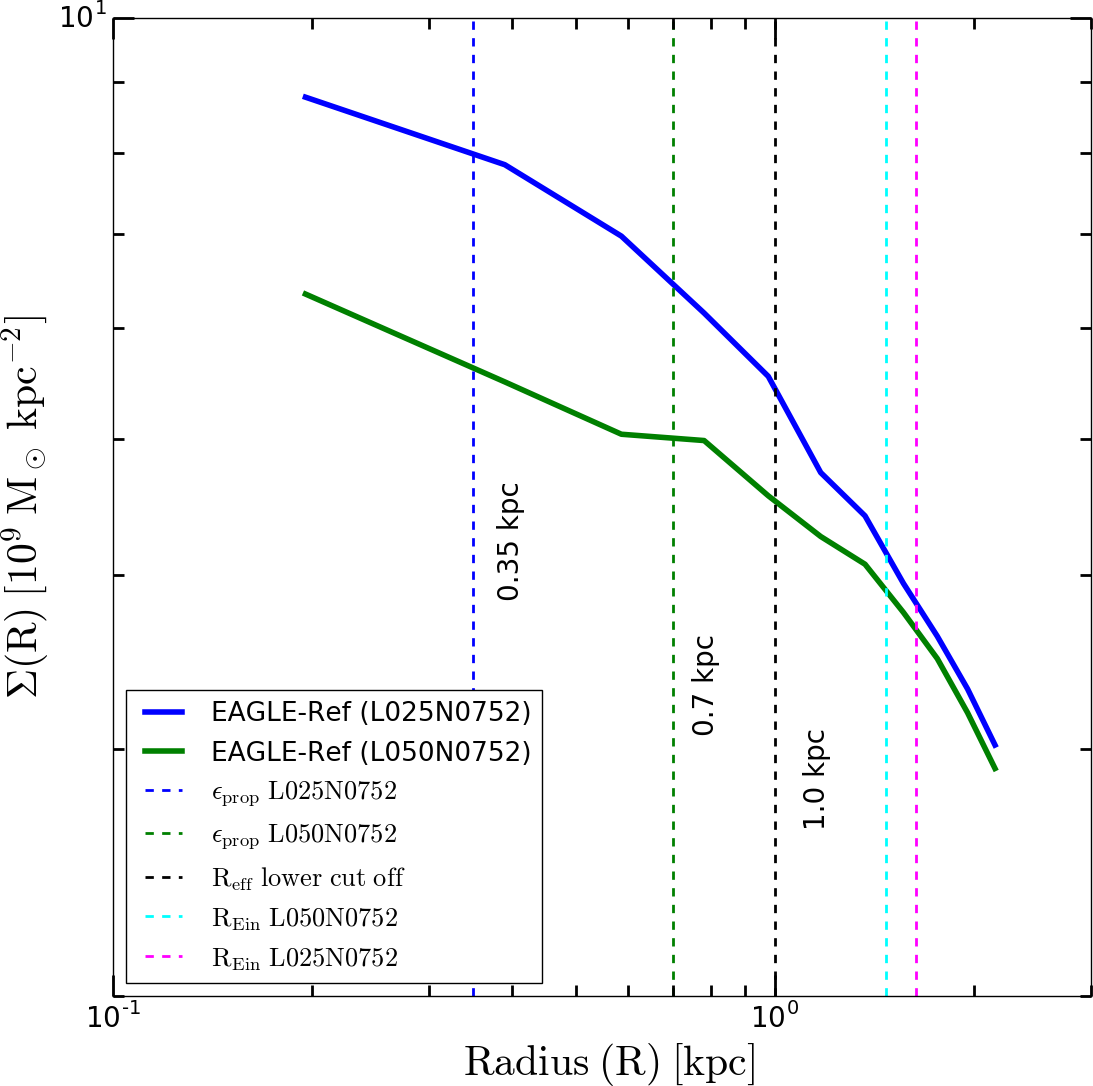}
\caption{Comparison of the surface mass density profile of two example ETGs extracted from the EAGLE-Reference runs L050N0752 and L025N0752 having similar stellar masses ($M_{\star}\; \sim \; 10^{10.6}\; M_{\bigodot}$). The effect of the smoothing kernel can be visualized when the slope flattens at their respective softening lengths. The Einstein radii 1.63 kpc and 1.47 kpc are also shown for the galaxy in L025N0752 (magenta dashed line) and L050N0752 (cyan dashed line) simulation respectively.}
\label{nm}
\end{figure}


{ Here we show how the spatial resolution affects the inner (< 1 kpc) region of EAGLE galaxies. We used two simulation boxes of EAGLE-Reference run i.e., L050N0752 and L025N0752 (the latter with higher resolution). In Figure~\ref{nm} we plot the surface mass density vs the radius for an example galaxy (after projection and lens creation). We can see that the slopes flatten at their respective softening lengths (represented by $ \epsilon_{\rm prop}$, as in Table~1). The radius where the two density profiles start to converge, is well inside the Einstein radii of these galaxies.\\

However, the effect of smoothing in the central region does not bias the strong lensing analysis, since we mask out the inner $7\times7$ pixels, which correspond to $1.4 \times 1.4$ kpc. Masking is a standard practice in observational analysis, too, where strong lenses are analysed after masking out the lensing galaxy. In order to not bias the results from simulation and to make an unbiased comparison with direct fitting results we perform this masking operation in our simulated galaxies (see Section 3). This aspect is very important as the cores can skew the density slopes obtained directly from simulations, if the mask is not used.

\section{Effect of source sizes and prior types}\label{Sapp}

We have used a sub sample of our simulated lenses to access the impact of source sizes and different prior settings. There lensing galaxies having $\rm M_{\star} >10^{11}M_{\bigodot}$ projected along 3 axes have been used. So lens-modeling results from a total of 9 lenses have been presented here. 

In Table~\ref{Sourceapp} we summarize the effect of different source sizes and prior types on the mean results of modeled density mass slope ($t_{\rm L}$) and shear components ($\gamma_{1}$ and $\gamma_{2}$). We used two different families of prior settings: Gaussian and uniform. The values of the priors (the mean $\mu $ and standard deviation $\sigma$ for gaussian priors and $minimum$ and $maximum$ of the range of values for uniform priors) are tabulated in Table~\ref{Sourceapp}. We find that there is no substantial effect of the priors on the final result. In this paper we used the Gaussian priors on $t_{\rm L}$ since the computational time is decreased by 30-40\% with respect to using uniform priors.


We also note that there is no significant improvement in the final results using more spatially extended sources. This is expected since with $R_{\rm eff}$=0.2 arcsec (typical SLACS source size; see \citealt{newton2011}) for an HST-ACS filter, the S/N is already sufficient to constrain lens parameters.
\begin{table}
\begin{center}
\caption{Comparison of the modeled density slopes and shear having different prior settings in {\tt LENSED} and using different source sizes. We note that the differences are minor and much smaller than the spread between systems or their typical errors ($t_{\rm L}$ is $\pm$ 0.05 and $\gamma_{\rm 1,2}$ is $\pm$ 0.001).}
\resizebox{\columnwidth}{!}{
  \begin{tabular}{ c c c c c c c}
    \hline
     & \multicolumn{3}{c}{EPL-Gaussian} & \multicolumn{3}{c}{EPL-Uniform} \\ 
     Prior on $t_L$& \multicolumn{3}{c}{$\mu$=1.1, $\sigma$=0.1} & \multicolumn{3}{c}{0 -- 2} \\
     Prior on $\gamma_1 \; \gamma_2$& \multicolumn{3}{c}{$\mu$=0.0, $\sigma$=0.01} & \multicolumn{3}{c}{-0.1 -- 0.1} \\\cdashline{1-1}
    Source & $t_{\rm L}$ & $\gamma_1$& $\gamma_2$ & $t_{\rm L}$ & $\gamma_1$&$\gamma_2$\\ \hline \hline
    ${}^{\dagger}$S\'{e}rsic, $R_{\rm eff}=0.2$ & 1.06 & -0.040&0.021  & 1.05 & -0.041&0.021\\ \hline
    ${}^{\dagger}$S\'{e}rsic, $R_{\rm eff}=0.3$ & 1.08 & -0.038&0.020 & 1.05 & -0.039&0.021 \\ \hline
    ${}^{\dagger}$S\'{e}rsic, $R_{\rm eff}=0.4$ & 1.06 & -0.036&0.020 & 1.04 & -0.040&0.021\\ \hline
    \multicolumn{7}{l}{$\dagger$ All other Source parameters have been kept same as Table \ref{priortable}}
  \end{tabular}
  }
  \label{Sourceapp}
\end{center}
\end{table}

\bsp	
\label{lastpage}
\end{document}